\documentclass[reprint,
superscriptaddress,
amsmath,amssymb,
aps,
prb,
]{revtex4-2}
\usepackage{graphicx}
\usepackage{dcolumn}
\usepackage{bm}
\usepackage{dsfont}
\usepackage{physics}
\usepackage{color}
\usepackage{soul}
\setstcolor{red}
\usepackage{dsfont}
\usepackage{ifthen}
\newboolean{showannotations}   
\setboolean{showannotations}{true} 
\newcommand*{\added}[1]{%
  \ifthenelse{\boolean{showannotations}}{{\color{blue}#1}}{#1}%
}

 \usepackage[thinc]{esdiff}

\newcommand*{\removed}[1]{%
  \ifthenelse{\boolean{showannotations}}{\st{#1}}{}%
}

\newcommand*{\change}[2]{%
  \ifthenelse{\boolean{showannotations}}{{\color{red}\st{#1}}{\color{blue}#2}}{#2}%
}

\begin{document} 

\title{Non-Markovian Quantum Dynamics of Exciton-Polaritons}
\author{Rajanya Sarkar}
\altaffiliation{These authors contributed equally to this work.}
\affiliation{Department of Chemistry, Texas A\&M University, College Station, Texas 77843, USA}

\author{Pritha Ghosh}%
\altaffiliation{These authors contributed equally to this work.}
\affiliation{Department of Chemistry, Texas A\&M University, College Station, Texas 77843, USA}

\author{Arshath Manjalingal}%
\affiliation{Department of Chemistry, Texas A\&M University, College Station, Texas 77843, USA}

\author{David R. Reichman}%
\email{drr2103@columbia.edu}
\affiliation{Department of Chemistry, Columbia University, New York, NY 10027, USA}

\author{Arkajit Mandal}%
\email{mandal@tamu.edu}
\affiliation{Department of Chemistry, Texas A\&M University, College Station, Texas 77843, USA}

\begin{abstract}
{\footnotesize Exciton-polaritons, hybrid light-matter quasiparticles formed when a material interacts with a confined electric field, have experimentally been shown to exhibit mesoscale coherent quantum propagation that remains robust at room temperature. However, an accurate and direct quantum dynamical simulation of this phenomenon that does not resort to semi-classical approximations is prohibitively expensive computationally, limiting the microscopic understanding of the rich dynamical interplay among phonons, photons, and electrons under collective light-matter coupling. To address this fundamental challenge, we develop a non-Markovian master equation approach which enables the fully quantum mechanical simulation of non-equilibrium exciton-polariton dynamics and captures phonon-induced decoherence and dissipation beyond the conventional Markovian limit. To carry out this task, a procedure is developed in which the wave vector space is coarse-grained and each diagonal element of the density matrix is evolved in parallel. To demonstrate the utility of this approach, we simulate exciton-polariton transport in TIPS-pentacene. We find that our approach reasonably captures the experimentally observed renormalization of the polariton group velocity, which originates from the phonon-induced non-Markovian Lamb shift. We further show that this renormalization cannot be reproduced within conventional Markovian theories. }
\end{abstract}

\maketitle
{\footnotesize
\textbf{Introduction.} In contrast to excitons, which typically exhibit incoherent diffusive transport at room temperature~\cite{sneyd2022new, mikhnenko2015exciton, tamai2015exciton, barua2026many}, exciton-polaritons, formed by coupling excitons to confined radiation modes inside an optical cavity, have experimentally been found to display coherent ballistic propagation over relatively long length and time scales~\cite{xu2023ultrafast, pandya2022tuning, balasubrahmaniyam2023enhanced, sandik2025cavity, hong2026exciton, berghuis2022controlling, chen2023unraveling}.  These observations suggest fundamentally new avenues for energy conversion, quantum sensing and transduction, opening up novel routes to tune the physicochemical properties of materials and molecular systems by leveraging the collective nature of light-matter interactions~\cite{sanvitto2016road, MandalCR2023, xiang2024molecular}. However, the fundamental microscopic principles governing the dynamics of exciton-polaritons remain under debate~\cite{fowler2026mapping, ying2025microscopic, blackham2025microscopic, berghuis2022controlling, xu2023ultrafast, balasubrahmaniyam2023enhanced, zhou2024nature, fitzgerald2025polariton, aroeira2025static, engelhardt2023polariton, tutunnikov2026noise} while a detailed and fully quantum mechanical simulation of their properties has remained computationally inaccessible. To enable simulation of exciton-polariton transport over relatively long length and time scales, recent theoretical investigations have resorted to semi-classical (or mixed quantum-classical) approximations wherein the phonon degrees of freedom are propagated classically~\cite{berghuis2022controlling, xu2023ultrafast, blackham2025microscopic,rahmanian2026exciton, sokolovskii2024one, sokolovskii2023multi, tichauer2023tuning, tichauer2021multi, chng2025quantum, ghosh2025mean, liu2025dissecting, poddar2026quantum, chng2026ab}. While numerically exact approaches based on hierarchical equations of motion~\cite{mitric2022spectral} or tensor networks~\cite{kloss2019multiset} have been used to resolve aspects of the dynamics of electrons or excitons coupled to phonons by exploiting the finite spatial extent of dynamically localized quasiparticles~\cite{citty2024mesohops, mohamed2026unraveling}, the direct extension to exciton-polariton transport remains computationally prohibitive in two and three spatial dimensions due to the low effective mass and relatively weak environmental coupling of exciton-polaritons,  which can remain delocalized and propagate coherently over long length and time scales. Existing quantum-mechanical treatments of phonon-induced dissipation of exciton-polariton propagation, meanwhile, have typically invoked a Markovian approximation~\cite{fowler2026mapping, fitzgerald2025polariton}, which we show here to be quantitatively inadequate, or restricted the dynamics to zero temperature~\cite{krupp2025quantum}, precluding a direct description of the room-temperature quantum dynamics.

In this work, we address this fundamental challenge by developing an efficient non-Markovian master equation approach that allows us to simulate the mesoscale quantum dynamical propagation of exciton-polaritons. Direct non-Markovian simulation is otherwise computationally prohibitive, as resolving two-dimensional exciton-polariton dynamics requires  $N\approx 10^8$ states, for which a naive implementation scales as $N^5$.   To make such simulations feasible, we (i) implement a time-local propagation scheme using a Prony decomposition of the memory kernel~\cite{UlrichJCP2004}, (ii) exploit the fact that each diagonal of the density matrix can be evaluated independently and therefore can be propagated in parallel, and (iii) introduce a coarse-grained representation of the memory kernel where k-space is divided into regions delineated by the curvature of the polariton band structure. The combination of these algorithmic techniques leads to an efficient dynamical scheme capable of capturing the dynamical loss of coherence in exciton-polariton wave packet propagation even in large systems over relatively long time scales.  We note that while here we treat the memory kernel to second order in the polariton-phonon coupling (which should be reasonable in most materials due to the reduced coupling of phonons to polaritons compared to excitons), our approach can be combined with exact numerical schemes for evaluating the memory function, such as those afforded by tensor network methods~\cite{kloss2019multiset}, or extended to higher order in the coupling~\cite{theta}. Using our method, we simulate organic exciton-polariton dynamics in TIPS-pentacene coupled to confined radiation modes, treating $N=12001 \times 12001 \sim 1.4\times10^8$ emitters with each coupled to a dissipative environment characterized microscopically by a spectral density obtained from on-the-fly DFTB/TD-DFTB simulations.~\cite{plotz2017spectral,Alkan2018}

We demonstrate that even with a high excitonic content, exciton-polaritons propagate ballistically for $\sim$100s of femtoseconds while renormalization of the polariton group velocity occurs due to interactions with phonons. We show that this effect originates from the imaginary part of the memory kernel, corresponding to a phonon-induced Lamb shift, and that its accurate evaluation requires a non-Markovian treatment. We find our results to be in reasonably good agreement with recent experiments, highlighting the predictive power of our quantum dynamical approach. Our work also demonstrates why a simple Redfield or Lindblad approach, as has been implemented in a recent work~\cite{fowler2026mapping}, will fail to capture this phonon induced effect on polariton propagation. 

\begin{figure*}
\centering
\includegraphics[width=1.0\linewidth]{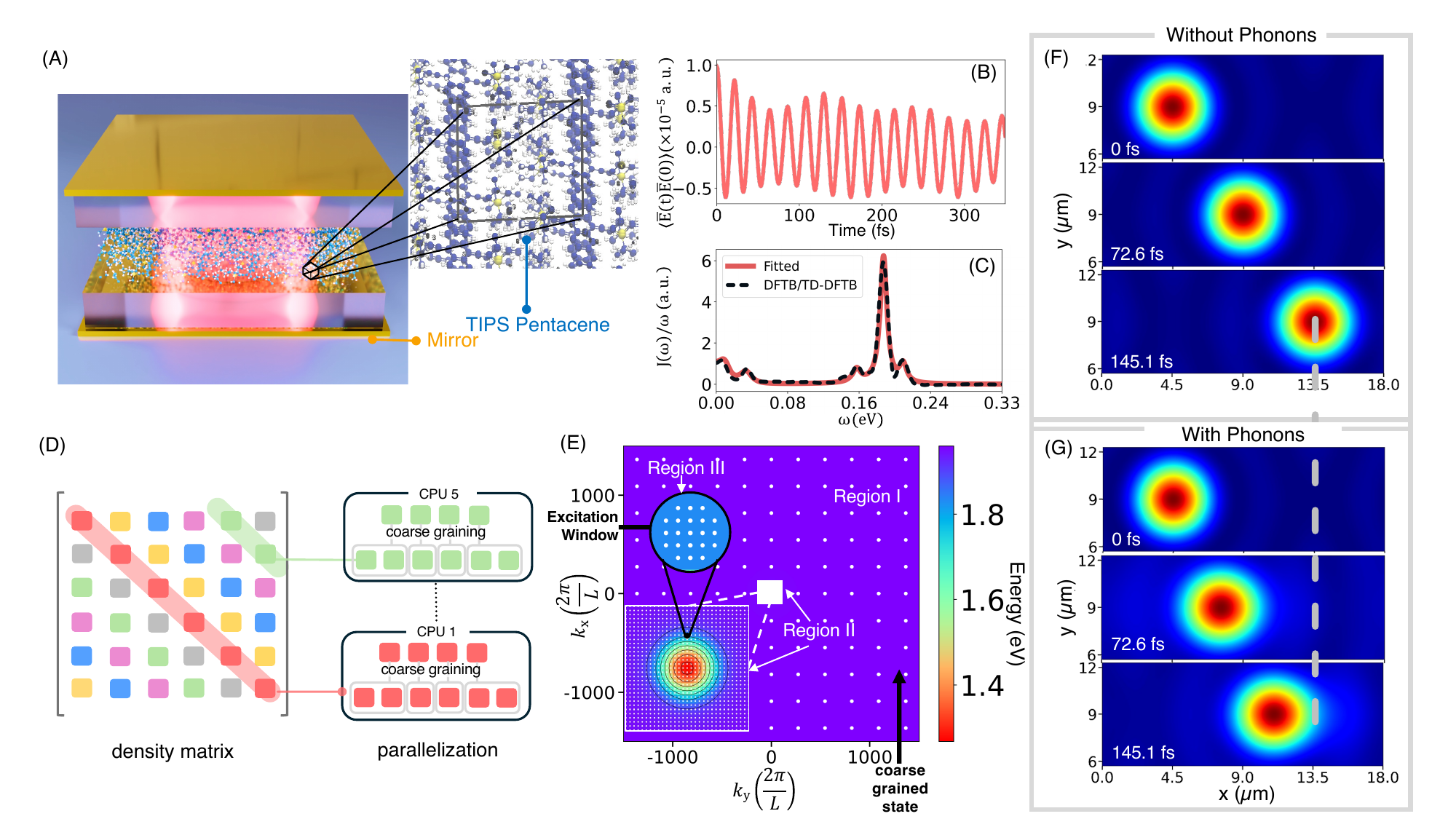}
\caption{\footnotesize (A) Schematic illustration of a Fabry-P\'{e}rot optical cavity filled with TIPS-pentacene. (B) Time-dependent energy gap correlation function computed using on-the-fly DFTB/TD-DFTB simulations. (C) Spectral density $J(\omega)$ obtained from the DFTB/TD-DFTB simulations compared with the fitted spectral density. (D) Schematic illustration of the parallel computational scheme for evolving the diagonal elements of the density matrix, which are subsequently coarse-grained. (E) Schematic illustration of our  coarse-graining approach. (F)-(G) Snapshots of the exciton-polariton propagation in two dimensional real space at 0 fs, 72.6 fs and 145.1 fs, respectively, in the absence (F) and presence (G) of phonons.}
\label{fig1}
\end{figure*}

\textbf{Theory.}  Following prior work~\cite{chng2025quantum, krupp2025quantum, KeelingARPC2020, haines2026mechanistic, aroeira2024coherent}, we consider a generalized multimode Holstein-Tavis-Cummings Hamiltonian, which describes an exciton-polariton system interacting with phonons, and is written as (using $\hbar = 1$ a.u.)
\begin{align}
\hat{H}_{\mathrm{LM}} &=  \sum_{\bf k} \bigg[\hat{X}_{\bf k}^{\dagger} \hat{X}_{\bf k} \varepsilon_{\bf k} +   g_c \big(\hat{a}^{\dagger}_{\bf k}\hat{X}_{\bf k}+ \hat{a}_{\bf k}\hat{X}_{\bf k}^{\dagger}\big)+  \hat{a}_{\bf k}^{\dagger}\hat{a}_{\bf k} \omega_{c}({\bf k})\bigg]\nonumber \\
&+ {\sum_{{\bf n},j} \frac{c_j}{\sqrt{2\omega_j}}  \hat{X}_{{\bf n}}^{\dagger} \hat{X}_{{\bf n}}  (\hat{b}^{\dagger}_{{\bf n},j} + \hat{b}_{{\bf n},j}) } + {\sum_{{\bf n}, j} \hat{b}^{\dagger}_{{\bf n},j}\hat{b}_{{\bf n},j}\omega_{j}},
\end{align} 
where $\hat X^\dagger_{\bf k} = \frac{1}{\sqrt{N}}\sum_{\bf n} e^{i {\bf k}\cdot {\bf R}_n } \hat X^\dagger_{\bf n}$ and  $\hat a^\dagger_{\bf k}$ are the excitonic and photonic creation operator of cavity mode, respectively, which are associated with the in-plane wave vector ${\bf k }= k_x \vec{x} + k_y \vec{y} $ with $\varepsilon_{\bf k}$ denoting the excitonic energy and $\omega_{c}({\bf k})$ the photon frequency. In this work, we consider a Fabry-P\'{e}rot optical cavity where the cavity photon frequency is written as $\omega_{c}({\bf k}) =  \sqrt{\frac{c^2}{\eta^2}(k_x^2 + k_y^2) + \omega_c^2(0)}$ where $\omega_c^2(0)$ is the quantized photon frequency at normal incidence with $c$ denoting speed of light and $\eta$ the refractive index of the material filling the optical cavity. Note that the cavity dispersion is much steeper than the excitonic dispersion; consequently, over the small-momentum range relevant to the exciton-polariton dynamics, the exciton dispersion can be treated as effectively flat, i.e., $\varepsilon_{\bf k} \approx \varepsilon_{\bf 0}$. 

The exciton-photon coupling strength is characterized by the constant $g_c$, and $\hat{b}_{{\bf n},j}^\dagger$ is the $j$th phonon creation operator (representing, e.g., the $j$th molecular mode) of frequency $\omega_j$ located at site $ {\bf n } = n_x \hat{x} + n_y \hat{y}$, with $c_j$ characterizing the strength of its coupling to the exciton. This exciton-phonon coupling, which we treat as independent of momentum, is described by the spectral density $J(\omega) =\frac{1}{2} \sum_{j} \frac{c_{j}^2}{\omega_{j}} \delta(\omega - \omega_{j}) = \sum_{\alpha} \frac{A_\alpha \omega }{[(\omega + \Omega_{\alpha})^2 + \Lambda_{\alpha}^2][(\omega - \Omega_{\alpha})^2 + \Lambda_{\alpha}^2]}$, where the parameters $\Omega_{\alpha}$, $\Lambda_{\alpha}$ and $A_\alpha$ are obtained by fitting the spectral density extracted from direct atomistic simulations (see Fig.~\ref{fig1}C), with further details provided in the supporting information (SI). In short, we obtain the spectral density by performing on-the-fly DFTB molecular dynamics simulations and then compute the energy correlation function  $\langle \bar{E}(t)\bar{E}(0) \rangle$ (see Fig.~\ref{fig1}B) using the TD-DFTB approach, where $\bar{E}(t) = \Delta E_{eg}(t) - \langle \Delta E_{eg}(t) \rangle$  and $\Delta E_{eg}(t)$ is the time-dependent ground to excited state energy gap.~\cite{gustin2023mapping, valleau2012alternatives, plotz2017spectral,makri1998quantum}  

To model the exciton-polariton dynamics relevant to recent experiments~\cite{xu2023ultrafast, hong2026exciton, balasubrahmaniyam2023enhanced}, we restrict the dynamics to the lower-polariton subspace by projecting the light--matter Hamiltonian as $\hat{H}_{\mathrm{LM}} \rightarrow    \mathcal{P}\hat{H}_{\mathrm{LM}}\mathcal{P}$, where $\mathcal{P} = \sum_{\bf k} \hat{P}_{\bf k}^{\dagger}\hat{P}_{\bf k}$ denotes the projector onto the lower-polariton manifold with $\hat{P}_{\bf k}^\dagger$ as the lower polariton creation operator. The lower polariton creation operator is written as $\hat{P}_{\bf k}^\dagger = \hat{X}_{\bf k}^{\dagger} \sin\phi_{\bf k}  + \hat{a}^{\dagger}_{\bf k} \cos \phi_{\bf k}$ where  $\phi_{\bf k} = \frac{1}{2} \tan^{-1}\big[\frac{2 g_c}{\omega_c({\bf k}) - \epsilon_0}\big]$ is the light-matter mixing angle. The light-matter Hamiltonian projected in the lower polariton subspace is written as

\begin{align}\label{LM}
\mathcal{P}\hat{H}_{\mathrm{LM}}\mathcal{P} &= \sum_{\bf k} \hat{P}_{\bf k}^{\dagger}\hat{P}_{\bf k} E_{\bf k}  + {\sum_{{\bf k}, j}  \hat{b}^{\dagger}_{{\bf k},j}\hat{b}_{{\bf k},j}\omega_{j}}  \\
&+  \sum_{{\bf k},{\bf \delta k}}  \hat{P}_{\bf k+\delta k}^{\dagger}\hat{P}_{{\bf k}} \Big[\gamma_{{\bf k} +\delta {\bf k},{\bf k}}\sum_j \frac{c_j}{\sqrt{2\omega_j}} (\hat{b}_{-\delta {\bf k}, j}^{\dagger} + \hat{b}_{\delta {\bf k},j})\Big],\nonumber
\end{align}

where $E_{\bf k} = \frac{1}{2}(\omega_c({\bf k}) + \epsilon_0) - \frac{1}{2}\sqrt{(\omega_c({\bf k}) - \epsilon_0)^2 + 4 g_c^2}$ is the lower polariton dispersion and the second line represents the polariton-phonon interaction term with $\gamma_{{\bf k} +\delta {\bf k},{\bf k}} = \gamma_{{\bf k} ,{\bf k}+\delta {\bf k}}= \frac{1}{\sqrt{N}}\sin\phi_{\bf k + \delta {\bf k}} \sin \phi_{\bf k}$. In the following, we simulate the exciton-polariton system by adopting an open quantum dynamics approach. Specifically, we start with the non-Markovian Nakajima-Zwanzig master equation~\cite{zwanzig1960ensemble, nakajima1958quantum, reineker1980exact} written as 
\begin{figure*}
\centering
\includegraphics[width=1.0\linewidth]{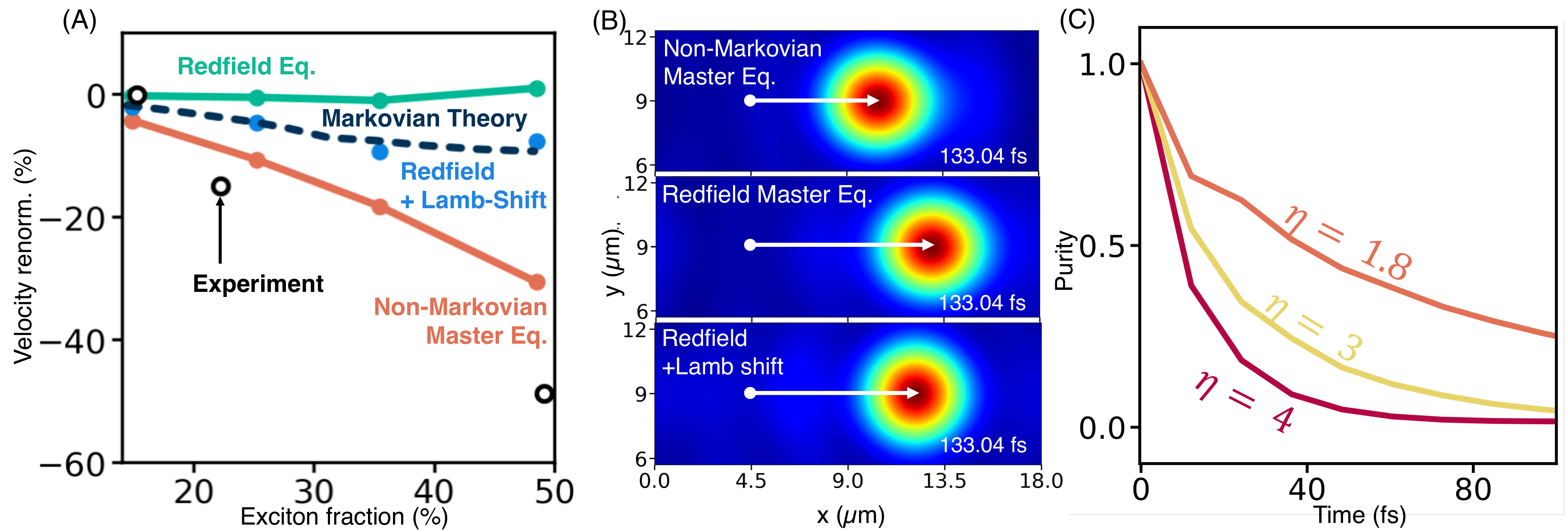}
\caption{\footnotesize  (A) Group velocities extracted from non-Markovian quantum dynamical simulations and various Markovian approaches,  compared to the experimentally determined group velocities obtained from Ref.~\cite{hong2026exciton}.  (B) A snapshot of the polaritonic density predicted from the non-Markovian master equation approach, the Redfield approach (with no Lamb shift term) and the Redfield approach with the Lamb shift term, at time t = 133.04 fs. (C) Time-dependent purity for three different refractive indices with $\eta = 1.8$ corresponding to the refractive index of TIPS-pentacene.}
\label{fig2}
\end{figure*}
\begin{equation}\label{NZ}
\dot{\rho}_{\bf kk'}(t) = -i \Delta E_{\bf k,\bf k'} \rho_{\bf kk'}(t) - \int_{0}^{t} d\tau \sum_{\bf qq'}\mathcal{K}_{\bf kk'}^{\bf qq'}(t-\tau)\rho_{\bf qq'}(\tau) ,
\end{equation}
where $\Delta E_{\bf k,\bf k'} = E_{\bf k} - E_{\bf k'}$ and  $\rho_{\bf kk'}(t)$ is the reduced density matrix element of the exciton-photon sub-system obtained after tracing over the phonon degrees of freedom. While an analytical expression for the exact memory kernel $\mathcal{K}_{\bf kk'}^{\bf qq'}(\tau)$ is unavailable, a tractable analytical expression correct to second order in the polariton-phonon coupling can be obtained, with the explicit expression provided in the SI. Directly simulating the mesoscale non-Markovian quantum dynamics of exciton-polariton using Eq.~\ref{NZ} is a formidable task given that the size of the exciton-photon Hilbert space ($N \sim 10^8$), with the memory kernel $\mathcal{K}_{\bf kk'}^{\bf qq'}(t-\tau)$ scaling as $N^4 \times N_t$, where $N_t$ denotes the number of time points used to resolve the memory time $\tau$. To overcome this challenge, we develop a time-local propagation approach which retains the non-Markovian nature of the dynamics which is combined with a non-uniform k-grid coarse-graining scheme that rapidly converges to the full dynamical evolution. To carry out this task we show that the equation-of-motion for the reduced density matrix written in the interaction picture may be expressed as
\begin{align}\label{CG-NMRE}
\dot{\rho}_{\bf k,k+\delta k}^{\mathrm{I}}(t) &=   \sum_{ {\bf q}\in \mathrm{CG}} n_{\bf q} \sum_\alpha\Big(\gamma_{\bf k, q}^2 I_{\bf k, q}^{{\delta \bf k},\alpha}(t) + \gamma_{\bf k + \delta k, q}^2 J_{\bf k, q}^{{\delta \bf k},\alpha}(t)  \Big)    \\
&+ \sum_{ {\bf q}\in \mathrm{CG}} n_{\bf q}\sum_\alpha\gamma_{\bf k+\delta k, q} \gamma_{\bf k , q+\delta k} \Big( F_{\bf k,q}^{{\delta \bf k},\alpha}(t) + G_{\bf k,q}^{{\delta \bf k},\alpha}(t)\Big),   \nonumber
\end{align}
where ${\rho}_{\bf k,k+\delta k}^{\mathrm{I}}(t) = e^{-i\Delta E_{\bf k, k+\delta k}t}{\rho}_{\bf k,k+\delta k}(t)$ is the reduced density matrix of the exciton-photon sub-system in the interaction picture and ${\bf q}\in \mathrm{CG}$ is a wave vector with the associated weight $n_{\bf q}$ in a particular region of k-space, as described further below. We arrive at this time-local propagation via a Prony decomposition of the bath correlation function~\cite{UlrichJCP2004}, i.e. $C(t) = \sum_{j} \frac{c_j^2}{2\omega_j}\Tr_{\mathrm{B}}[\hat{b}_{{\bf k},j}^\dagger(t) \hat{b}_{{\bf k},j}(0) \hat{\rho}_{\mathrm{B}}^{eq}] = \sum_\alpha {\mathcal B}_\alpha e^{-\xi_\alpha t}$, where $\hat{\rho}_{\mathrm{B}}^{eq}$ is equilibrium density matrix of the phonon degrees of freedom, with the explicit expressions for ${\mathcal B}_\alpha$ and $\xi_\alpha$ provided in the SI for the fitted spectral density (see Fig.~\ref{fig1}C) obtained from DFTB/TD-DFTB simulations. We introduce four time-dependent auxiliary tensors, $I_{\bf k, q}^{{\delta \bf k},\alpha}(t)$, $J_{\bf k, q}^{{\delta \bf k},\alpha}(t)$,  $F_{\bf k, q}^{{\delta \bf k},\alpha}(t)$ and $G_{\bf k, q}^{{\delta \bf k},\alpha}(t)$, which are evolved along with the density matrix ${\rho}_{\bf k,k+\delta k}^{\mathrm{I}}(t)$. These auxiliary tensors are propagated as

\begin{align}
I_{\bf k, q}^{{\delta \bf k},\alpha}(t + \delta t) &= I_{\bf k, q}^{{\delta \bf k},\alpha}(t)  e^{(i\Delta E_{\bf k, q} - \xi_\alpha)\delta t}  \\
&- {\rho}_{\bf k,k+\delta k}^{\mathrm{I}}(t)\frac{{\mathcal{B}_{\alpha}}   \Big[1- e^{(i\Delta E_{\bf k, q}-\xi_\alpha) \delta t} \Big]}{i \Delta E_{\bf k, q} -\xi_\alpha }, \nonumber \\
J_{\bf k, q}^{{\delta \bf k},\alpha}(t + \delta t) &= J_{\bf k, q}^{{\delta \bf k},\alpha}(t)  e^{(i\Delta E_{\bf q, k+\delta k}-\xi_\alpha^*)\delta t}  \\
&- {\rho}_{\bf k,k+\delta k}^{\mathrm{I}}(t)\frac{{\mathcal{B}_{\alpha}^*}   \Big[1- e^{(i\Delta E_{\bf q, k+\delta k}-\xi_\alpha^*)\delta t}\Big]}{i\Delta E_{\bf q, k+\delta k}-\xi_\alpha^* }, \nonumber \\
G_{\bf k, q}^{{\delta \bf k},\alpha}(t + \delta t) &=  G_{\bf k, q}^{{\delta \bf k},\alpha}(t) e^{(i\Delta E_{\bf q + \delta k, k+\delta k}-\xi_\alpha)\delta t}  \\
&- {\rho}_{\bf q,q+\delta k}^{\mathrm{I}}(t) \frac{\mathcal{B}_{\alpha}e^{i\Delta E_{\bf q + \delta k, k+\delta k} (t + \delta t)} }{i\Delta E_{\bf q, k} - \xi_\alpha}\nonumber\\
&\times \bigg[e^{-i\Delta E_{\bf q, k}(t + \delta t)} - e^{-i \Delta E_{\bf q, k}t -\xi_\alpha \delta t } \bigg], \nonumber \\
F_{\bf k, q}^{{\delta \bf k},\alpha}(t + \delta t) &= F_{\bf k, q}^{{\delta \bf k},\alpha}(t) e^{(i\Delta E_{\bf k , q}-\xi_\alpha^*)\delta t}  \\
&-  {\rho}_{\bf q,q+\delta k}^{\mathrm{I}}(t) \frac{\mathcal{B}_{\alpha}^* e^{i\Delta E_{\bf  k, q} (t + \delta t)} }{i\Delta E_{\bf k+\delta k, q+\delta k} - \xi_\alpha^*}  \nonumber \\
&\times \bigg[ e^{-i\Delta E_{\bf k + \delta k, q + \delta k}(t + \delta t)} - e^{-i \Delta E_{\bf k + \delta k, q + \delta k}t -\xi_\alpha^* \delta t } \bigg]. \nonumber
\end{align}

Note that each diagonal element of the reduced density matrix corresponding to $\delta k$ can be simulated in parallel (illustrated in Fig.~\ref{fig1}D) which vastly reduces the overhead of the computation. For each $\delta k$-th diagonal of the density matrix, the summation on the right hand side of the equation is coarse-grained, such that the wave vectors ${\bf q}$ under the summation in Eq.~\ref{CG-NMRE} are sampled with $n_{\bf q}$ as the associated weights with $N = \sum_{\bf q\in \mathrm{CG}} n_{\bf q}$. To sample the wave vectors we use a three-tiered approach with the sampling algorithm dividing the momentum space into three resolution regimes as indicated in Fig.~\ref{fig1}E. Region I corresponds to flat regions of the dispersion, which we coarse-grain heavily by sparsely sampling the wave vector space. In contrast, region II of the dispersion  which lies at the center of the Brillouin zone, is coarse-grained to a lesser extent. Finally, region III, which represents the region of our initial excitation, is not coarse-grained, such that $n_{\bf q}=1$ throughout this region.

{\bf Results and Discussion.} 
In Fig.~\ref{fig1}F, we present the quantum dynamical propagation of the exciton polariton density in the absence of coupling to phonons, i.e. $c_j = 0$. We prepare our initial state as $|\psi(0)\rangle = \sum_{\bf k\in Region ~III} e^{-i{\bf k}\cdot {\bf R_0}}\hat{P}_{\bf k}^{\dagger} |0\rangle $ (with $\hat{\rho}(t) = |\psi(0)\rangle\langle \psi(0)|$), which is a superposition of polaritonic states in Region III, with the density centered at ${\bf R_0} = [4.5 ~\mu m, 9.0 ~\mu m]$ in real space. In the absence of phonons, we observe simple ballistic propagation along the $x$ direction, as our initial state has a finite momentum along $x$. Importantly, upon including polariton-phonon interactions in Fig.~\ref{fig1}G, we find that the polaritonic density continues to propagate ballistically, albeit with a clearly reduced velocity of the {\em wave front}.  This renormalization of group velocity is consistent with recent experimental observations~\cite{xu2023ultrafast, pandya2022tuning, balasubrahmaniyam2023enhanced, sandik2025cavity, hong2026exciton}.   Interestingly, although not the focus here, it should be noted that non-Markovian effects can even play a role in renormalizing transport coefficients.  For example, within standard Markovian theories of exciton diffusion ~\cite{10.1063/1.434789,Munn01011980,Cheng2008}, a second-order treatment of the coupling to phonons can be shown analytically to lead to the standard Boltzmann expression for the diffusion constant.  However, this result, which depends on the equilibrium density matrix, band velocities, and scattering rates, is strictly valid only in the long-time limit, where the system's density matrix evolves on a timescale much slower than the bath relaxation time, namely, in the strict Markovian limit. In exciton-polariton systems, however, cavity loss restricts the overall dynamics to only hundreds of femtoseconds. Consequently, even after the exciton-polariton has fully decohered and a quasi-diffusive regime has been reached, the observed diffusion constant  differs drastically from its equilibrium value because the density matrix has not yet fully equilibrated. This aspect has been discussed in the SI.

In Fig.~\ref{fig2}A, we compute the $\%$ velocity renormalization in the TIPS-pentacene coupled cavity setup,  defined as $\frac{1}{v_g^{0}} (v_g -  v_g^{0})\times 100$ where $v_g$ is the observed group velocity in the presence of phonons while $v_g^0$ is the group velocity in the absence of the phonons.  Fig.~\ref{fig2}A demonstrates that the group velocity renormalization obtained using our non-Markovian approach is in reasonable agreement with experimental results digitized from a recent study~\cite{hong2026exciton}. We note that in Ref.~\cite{hong2026exciton} a three-band model was used to compute the excitonic character, in contrast to the two band model used here. However, in the SI we show that the additional matter state in the three-band model does not alter the excitonic character of the lower polariton to any non-negligible extent.  The lack of full quantitative agreement with experiment likely originates from some combination of inaccuracies in the parametrization of the model Hamiltonian and the second-order treatment of the memory function.  Both of these aspects could be systematically improved.

We compare our results to those simulated via the Redfield equation which is obtained by making the Markovian approximation and ignoring the Lamb shift, i.e., $\int_{0}^{t} d\tau \sum_{\bf qq'}\mathcal{K}_{\bf kk'}^{\bf qq'}(t-\tau)\rho_{\bf qq'}(\tau) \rightarrow \sum_{\bf qq'}\rho_{\bf qq'}(t)~\mathrm{Re}\Big[\int_{0}^{\infty} d\tau \mathcal{K}_{\bf kk'}^{\bf qq'}(\tau) \Big]$, such that the reduced density matrix is evolved as
\begin{align}\label{Markov}
\dot{\rho}_{\bf kk'}(t) = -i \Delta E_{\bf k,\bf k'} \rho_{\bf kk'}(t) - \sum_{\bf qq'}  \mathcal{R}_{\bf kk'}^{\bf qq'}  \rho_{\bf qq'}(t),
\end{align}
where $\mathcal{R}_{\bf kk'}^{\bf qq'} = \mathrm{Re}\Big[\int_{0}^{\infty} d\tau \mathcal{K}_{\bf kk'}^{\bf qq'}(\tau) \Big]$ is the Redfield tensor. Details of our implementation are provided in the supporting information. 

Fig.~\ref{fig2}A clearly demonstrates that such an approach, or related variants (e.g. the Lindblad approach), fail to capture this renormalization of group velocity. The failure of the Redfield theory (when ignoring the Lamb shift) can be more prominently observed in Fig.~\ref{fig2}B which displays the polariton density at $t  \sim 133$ fs. Note that, in addition to exhibiting no suppression of the group velocity, the spatial profile of the polaritonic density is also markedly different in the Redfield case. Overall, despite the widespread use of Redfield or Lindblad approaches for incorporating phonon induced dissipation~\cite{fowler2026mapping, fitzgerald2025polariton, osipov2023transport, son2022energy, saez2018organic,zhang2018monitoring, climent2022not}, our work demonstrates the limitations of using these approaches for simulating exciton-polariton dynamics. The key reason why this Redfield approach fails in this regard is the lack of the Lamb shift. When adding back the Lamb shift, $\mathcal{R}_{\bf kk'}^{\bf qq'} \rightarrow  \int_{0}^{\infty} d\tau \mathcal{K}_{\bf kk'}^{\bf qq'}(\tau)$, the phonon-induced renormalization of the polariton group velocity is partially recovered. Within the Markovian limit, the group velocity renormalization for an excitation at $E_{\bf k_0}$ can also be provided analytically (see the dashed solid line in Fig.~\ref{fig2}A) as
\begin{align}
 \frac{v_g -  v_g^{0}}{v_g^{0}}  = -    \frac{\sum_{\alpha, {\bf p}} \mathrm{Im}\Bigg[ \frac{{\mathcal{B}}_\alpha     \gamma_{\bf \footnotesize k_0,p}^2}{\xi_{\alpha} -i (E_{\bf k_0 + 
\delta k}- E_{\bf p})}    +       \frac{ {\mathcal{B}}^*_\alpha   \gamma_{\bf k_0+\delta k,p}^2}{ \xi^*_{\alpha}- i (E_{\bf p}- E_{\bf k_0})} \Bigg] }{E_{\bf k_0 + 
\delta k}- E_{\bf k}},
\end{align}
where $\delta k \rightarrow 0$. 
However, the predicted velocity renormalization (within the Markovian limit)  substantially deviates from the experimental results, illustrating the need for a non-Markovian treatment.

To understand the origin of the relatively long-lived coherent nature of exciton-polariton propagation within our framework,  we compute the time-dependent purity, defined as $\mathrm{Tr}[\hat{\rho}^2(t)]$, which quantifies phonon-induced decoherence. Note that computing the time-dependent purity by first constructing the density matrix is a computationally challenging task given the size of the system considered here. Instead, we compute the purity by noting that 
\begin{align}
    \mathrm{Tr}[\hat{\rho}^2(t)] = \sum_{\bf k, \delta k}|\rho_{\bf k, k+\delta k}(t)|^2  \approx \sum_{\bf \delta k }  \sum_{\bf k\in \mathrm{CG}} n_{\bf k}^2|\rho_{\bf k, k+\delta k}(t)|^2 
\end{align} 

where each term within the outer summation, $\sum_{\bf k \in \mathrm{CG}} n_{\bf k}^2|\rho_{\bf k, k+\delta k}|^2 $, is computed in parallel within our coarse-graining approach. Our analysis reveals that one of the key reasons for the long-lived exciton-polariton coherence is the lower density of states associated with the polariton dispersion compared to a bare excitonic dispersion. In particular, phonon-induced intraband scattering is a key pathway for decoherence in such systems, and a lower density of available states consequently suppresses the phonon-induced scattering rate. To test this hypothesis, we simply increase the refractive index $\eta$ in our simulations which effectively renders the photonic band (and consequently the polaritonic band)
much flatter, thereby increasing the density of states. Fig.~\ref{fig2}C presents the time-dependent purity for three different refractive indices. Note that, to make a fair comparison, we consider an excitation window with the same excitonic character of $\sim 48\%$. Clearly, increasing the refractive index increases phonon induced decoherence which leads to a sharper decrease in purity, thereby supporting this reasoning.

{\bf Summary and Conclusion.} In summary, we have developed a highly efficient non-Markovian master equation approach which opens the door to the simulation of quantum dynamics of exciton-polaritons in two and higher dimensions. This is enabled by expressing the memory function equation in a time-local form combined with a coarse-graining scheme which non-uniformly discretizes k-space into regions determined by the steepness of the polariton band structure. Together with the efficient parallel propagation of the 
resulting auxiliary tensors and the reduced density matrix elements, exciton-polariton transport over long length and time scales can be reliably performed. We combine our approach with parameters obtained from DFTB/TD-DFTB simulations to simulate exciton-polariton transport in TIPS-pentacene coupled to an optical cavity. This approach is shown to be in good agreement with recent experimental work~\cite{hong2026exciton}. Our work also sheds light on the limitations of Markovian approaches to exciton-polariton transport, and provides a new perspective on the long-lived nature of the exciton-polariton coherence in solids. 

{\bf Acknowledgments}. 
This work was supported by  the U.S. National Science Foundation (NSF) under Grant
No. CHE-2611431 and partially by the Texas A\&M startup funds. This work used TAMU ACES and LAUNCH clusters at the Texas A\&M University through allocation  PHY230021 from the Advanced Cyberinfrastructure Coordination Ecosystem: Services \& Support (ACCESS) program, which is supported by National Science Foundation grants \#2138259, \#2138286, \#2138307, \#2137603, and \#2138296.  The authors appreciate discussions with Logan Blackham, Saeed Rahmanian Koshkaki, Sachith Wickramasinghe, Amir Amini and Michael Fowler.

}
\bibliography{bib.bib}
\bibliographystyle{naturemag}

\end{document}


\title{Supplementary Information for Non-Markovian Quantum Dynamics of Exciton-Polaritons }
\author{Rajanya Sarkar}
\altaffiliation{These authors contributed equally to this work.}
\affiliation{Department of Chemistry, Texas A\&M University, College Station, Texas 77843, USA}

\author{Pritha Ghosh}%
\altaffiliation{These authors contributed equally to this work.}
\affiliation{Department of Chemistry, Texas A\&M University, College Station, Texas 77843, USA}

\author{Arshath Manjalingal}%
\affiliation{Department of Chemistry, Texas A\&M University, College Station, Texas 77843, USA}

\author{David R. Reichman}%
\email{drr2103@columbia.edu}
\affiliation{Department of Chemistry, Columbia University, New York, NY 10027, USA}

\author{Arkajit Mandal}%
\email{mandal@tamu.edu}
\affiliation{Department of Chemistry, Texas A\&M University, College Station, Texas 77843, USA}
\maketitle

\section{Exciton-Polariton Hamiltonian}
The light–matter Hamiltonian considered in the main-text can be re-expressed in the form $\hat{H}_{\mathrm{LM}}=\hat{H}_S+\hat{H}_B+\hat{H}_I$, 
where $\hat{H}_S$  describes a system Hamiltonian, $\hat{H}_B$ is a bath Hamiltonian and $\hat{V}$ describes their interactions. Specifically, $\hat{H}_S$ describes the exciton-photon part of the Hamiltonian which is written as
\begin{equation}
  \hat{H}_S =  \sum_{\bf k} \bigg[\hat{X}_{\bf k}^{\dagger} \hat{X}_{\bf k} \varepsilon_{\bf k} +   g_c \big(\hat{a}^{\dagger}_{\bf k}\hat{X}_{\bf k}+ \hat{a}_{\bf k}\hat{X}_{\bf k}^{\dagger}\big)+  \hat{a}_{\bf k}^{\dagger}\hat{a}_{\bf k} \omega_{c}({\bf k})\bigg],
\end{equation}
where $\hat X^\dagger_{\bf k} = \frac{1}{\sqrt{N}}\sum_{\bf n} e^{i {\bf k}\cdot {\bf R}_n } \hat X^\dagger_{\bf n}$ and  $\hat a^\dagger_{\bf k}$ are the excitonic and photonic creation operator of cavity mode, respectively, which are associated with the in-plane wavevector ${\bf k }= k_x \vec{x} + k_y \vec{y} $ with $\varepsilon_{\bf k}$ denoting the excitonic energy and $\omega_{c}({\bf k})$ the photon frequency. Meanwhile, the bath Hamiltonian $\hat{H}_B$ describes a collection of phonon modes which is written as
\begin{equation}
    \hat{H}_B= {\sum_{{\bf n}, j} \hat{b}^{\dagger}_{{\bf n},j}\hat{b}_{{\bf n},j}\omega_{j}},
\end{equation}
where $\hat{b}^{\dagger}_{{\bf n},j}$ is the phonon creation operator associated with the phonon mode $j$ at site ${\bf n}$. Finally, the interaction term assumes a typical Holstein form~\cite{wang2011mixed, krotz2021reciprocal, krotz2022reciprocal, blackham2025microscopic} and is written as
\begin{equation}
   \hat{H}_I= {\sum_{{\bf n},j} \frac{c_j}{\sqrt{2\omega_j}}  \hat{X}_{{\bf n}}^{\dagger} \hat{X}_{{\bf n}}  (\hat{b}^{\dagger}_{{\bf n},j} + \hat{b}_{{\bf n},j}) } ,
\end{equation}
where  phonon frequency $\omega_j$ and exciton-phonon coupling $c_j$ are described by the spectral density $J(\omega) = \sum_{j} \frac{c_{j}^2}{2\omega_{j}} \delta(\omega - \omega_{j})$. We perform our quantum dynamical simulation by projecting the total Hamiltonian within the lower polariton subspace, where the projected Hamiltonian written as 
\begin{align}\label{LM}
\hat{\mathcal{H}} &=  \mathcal{P}(\hat{{H}}_S +\hat{{H}}_B + \hat{H}_I)\mathcal{P} =  \hat{\mathcal{H}}_S +\hat{\mathcal{H}}_B + \hat{V}  \\
&= \sum_{\bf k} \hat{P}_{\bf k}^{\dagger}\hat{P}_{\bf k} E_{\bf k}  + {\sum_{{\bf k}, j}  \hat{b}^{\dagger}_{{\bf k},j}\hat{b}_{{\bf k},j}\omega_{j}}  +  \sum_{{\bf k},{\bf \delta k}}  \hat{P}_{\bf k+\delta k}^{\dagger}\hat{P}_{{\bf k}} \Big[\gamma_{{\bf k} +\delta {\bf k},{\bf k}}\sum_j \frac{c_j}{\sqrt{2\omega_j}} (\hat{b}_{-\delta {\bf k}, j}^{\dagger} + \hat{b}_{\delta {\bf k},j})\Big],
\end{align}
where $\mathcal{P} = \sum_{\bf k} \hat{P}_{\bf k}^{\dagger}\hat{P}_{\bf k}$ is the projection operator, $\hat{P}_{\bf k}^{\dagger}$ as the lower polariton creation operator, $\hat{b}_{\delta {\bf k}, j}^{\dagger} = \frac{1}{\sqrt{N}}\sum_{{\bf n}}\hat{b}_{ {\bf n}, j}^{\dagger}e^{i\delta{\bf k}\cdot {\bf n}}$ is the phonon creation operator in the reciprocal space with $N$ as the total number of excitonic sites, and $\gamma_{{\bf k} +\delta {\bf k},{\bf k}}$ is renormalization factor for the polaritonic-phonon coupling. The lower polariton operator  is written as $\hat{P}_{\bf k}^\dagger = \hat{X}_{\bf k}^{\dagger} \sin\phi_{\bf k}  + \hat{a}^{\dagger}_{\bf k} \cos \phi_{\bf k}$ where  $\phi_{\bf k} = \frac{1}{2} \tan^{-1}\big[\frac{2 g_c}{\omega_c({\bf k}) - \epsilon_0}\big]$ is the light-matter mixing angle. The renormalization factor for the polaritonic-phonon coupling terms is written as $\gamma_{{\bf k} +\delta {\bf k},{\bf k}} = \frac{1}{\sqrt{N}} \sin\phi_{\bf k}\sin\phi_{\bf k}$.  \\\\\\\\

\section{Non-Markovian Master Equation for Exciton-Polaritons}
We perform quantum dynamical simulations using the polariton-phonon Hamiltonian expressed in Eq.~\ref{LM}. Specifically, we evolve the reduced density matrix $\hat{\rho} (t)= \mathrm{Tr}_B[\hat{\rho}_{S+B} (t)]$ of the exciton-polariton (sub) system, where $\hat{\rho}_{S+B} (t)$ is the full density matrix and $\mathrm{Tr}_B[...]$ indicates trace with respect to the phonon bath. The general non-Markovian master equation is written as
\begin{align}
\dot{\hat{\rho}}(t) = -i \mathcal{L_S} \hat{\rho}(t) - \int_0^t \mathcal{K}(t - \tau) \hat{\rho}(\tau) d\tau,
\end{align}
where $\mathcal{L}_\mathcal{S}$ is the system Liouvillian, such that $\mathcal{L}_\mathcal{S}  \hat{\rho}(t) = [\hat{\mathcal{H}}_{S}, \hat{\rho}(t)]$ and $\mathcal{K}(t - \tau)$ is the bath memory kernel. We introduce the interaction Liouvillian, $\mathcal{L'} = \mathcal{L}-\mathcal{L}_0$ where $\mathcal{L} =[\hat{\mathcal{H}}, ... ]$  and $\mathcal{L}_0 =[\hat{\mathcal{H}}_S + \hat{\mathcal{H}}_B, ... ]$. A tractable expression for the bath memory kernel can be written (up to the second order in the interaction Liouvillian $\mathcal{L'}$) as

\begin{align} \label{memory-kernel}
\mathcal{K}(t) \approx \operatorname{Tr}_B[\mathcal{L'} e^{-i \mathcal{L}_0 t} \mathcal{L'} \rho_B^\mathrm{eq} \ldots].
\end{align}

The explicit form of the non-Markovian master equation using the memory-kernel in Eq.~\ref{memory-kernel}, using the polariton basis $\{\hat{P}_{\bf k}^{\dagger}|{\bf 0}\rangle\} \equiv \{|{\bf k}\rangle\} $ is written as 

\begin{align}
\dot{\rho}_{\bf kk'}(t) =& -i \langle {\bf k}|\mathcal{L_S} \hat{\rho}(t)|{\bf k'} \rangle  - \int_0^t \langle {\bf k}|\mathcal{K}(t - \tau) \hat{\rho}(\tau)|{\bf k'} \rangle d\tau\nonumber\\
=& -i(E_{\bf k}-E_{\bf k'})\rho_{\bf kk'}(t)- \int_0^t \langle {\bf k}|\mathcal{K}(t - \tau) \hat{\rho}(\tau)|{\bf k'} \rangle d\tau.
\end{align}

Specifically the term within the time integral  is explicitly written as   

\begin{align}
\langle {\bf k}|\mathcal{K}(t-\tau)\hat{\rho}(\tau)|{\bf k'}\rangle = 
\langle {\bf k}|\operatorname{Tr}_B [\mathcal{L}' e^{-i L_0 (t-\tau)} \mathcal{L}' \hat{\rho}_B^\mathrm{eq} \hat{\rho}(\tau)] |{\bf k'}\rangle
= \langle {\bf k}| \operatorname{Tr}_B [\mathcal{L}' e^{-i L_0 (t-\tau)} [\hat{V},\hat{\rho}(\tau) \hat{\rho}_B^\mathrm{eq}]] |{\bf k'}\rangle,
\end{align}
which can be decomposed as $\langle {\bf k}|\mathcal{K}(t-\tau)\hat{\rho}(\tau)|{\bf k'}\rangle = \mathrm I - \mathrm {II}- \mathrm {III} + \mathrm{IV}$ with each of the terms $\mathrm {I}- \mathrm {IV}$ written as 
\begin{align}
\mathrm{I} &= \langle {\bf k}| \operatorname{Tr}_B [\hat{V} \hat{U}_{t-\tau} \hat{V}\hat{\rho}(\tau)\hat{\rho}_B^\mathrm{eq} \hat{U}_{t-\tau}^\dagger] |{\bf k'} \rangle ,\\
\mathrm{II} &= \langle {\bf k}| \operatorname{Tr}_B [\hat{V} \hat{U}_{t-\tau} \hat{\rho}(\tau)\hat{\rho}_B^\mathrm{eq} \hat{V}\hat{U}_{t-\tau}^\dagger] |{\bf k'}\rangle ,\\
\mathrm{III} &= \langle {\bf k}| \operatorname{Tr}_B [\hat{U}_{t-\tau} \hat{V} \hat{\rho}(\tau)\hat{\rho}_B^\mathrm{eq} \hat{U}_{t-\tau}^\dagger \hat{V}]|{\bf k'}\rangle, \\
\mathrm{IV} &= \langle {\bf k}| \operatorname{Tr}_B [\hat{U}_{t-\tau} \hat{\rho}(\tau)\hat{\rho}_B^\mathrm{eq} \hat{V} \hat{U}_{t-\tau}^\dagger \hat{V}]|{\bf k'}\rangle.
\end{align}
Here, $\hat{\rho}_B^\mathrm{eq} \propto e^{-\beta \hat{\mathcal{H}}_B}$ is the equilibrium density matrix of the bath and $\hat{U}_{t-\tau} = e^{-i (\hat{\mathcal{H}}_S + \hat{\mathcal{H}}_B)(t-\tau)} = \hat{U}_S({t-\tau}) \hat{U}_B({t-\tau})$. Below, we write the explicit form of the terms $\mathrm {I}- \mathrm {IV}$  as (using $\tilde{t} = t-\tau$ and  ${\bf k'=k+\delta k}$)
\begin{align} 
\mathrm{I} = &\sum_{\mathbf{p},\mathbf{p'}}  \mathrm{Tr}_B\bigg[ \langle \mathbf{k} |\hat{V}| \mathbf{p} \rangle \cdot \langle  \mathbf{p} |\hat{U}_{B}(\tilde{t}) \hat{U}_{S}(\tilde{t}) \hat{V} | \mathbf{p'} \rangle \cdot  \langle  \mathbf{p'}|\hat{\rho}(\tau)| \mathbf{k} +\delta \mathbf{k} \rangle \cdot  \hat{\rho}_B^{eq}\hat{U}^{\dagger}_{B}(\tilde{t})\bigg]e^{iE_{\bf k} \tilde{t}} \nonumber\\
  =&\sum_{\bf p,\mathbf{p'}}  \mathrm{Tr}_B\bigg[ \hat{U}^{\dagger}_{B}(\tilde{t})  {V}_{\bf k ,\bf p} \hat{U}_{B}(\tilde{t}) \cdot  {V}_{\bf p  , \bf p' }\cdot  \rho_{\bf p' , \bf k +\delta k}(\tau) \cdot  \hat{\rho}_B^{eq}\bigg]e^{i (E_{\bf k +\delta k}- E_{\bf p}) \tilde{t}} \nonumber\\
    =&\sum_{\bf p, {\bf p'}}  \mathrm{Tr}_B\bigg[  V_{\bf k,\bf p}(\tilde{t}) \cdot  V_{\bf p, \bf p'}\cdot  \rho_B^{eq}\bigg]\cdot  \rho_{\bf p', \bf k +\delta \bf k}(\tau)e^{i (E_{\bf k +\delta k}- E_{\bf p}) \tilde{t}}  \nonumber \\
    =&\sum_{\bf p, {\bf p'}}  \Big\langle V_{\bf k, \bf p}(\tilde{t}) \cdot V_{\bf p, \bf p'}(0)   \Big\rangle\cdot  \rho_{\bf p', \bf k +\delta \bf k}(\tau)e^{i (E_{\bf k +\delta k}- E_{\bf p}) \tilde{t}}. \label{term-I} \\
\mathrm{II} = & \sum_{\bf p, {\bf p'}}  \mathrm{Tr}_{\mathrm{B}} \left[\big \langle {\bf k} | \hat{V} | {\bf p}\big \rangle \big\langle {\bf p}|\hat{U}_{\tilde{t}} \hat{\rho}(\tau) \hat{\rho}_B^{eq} | {\bf p'}\big \rangle \big\langle {\bf p'}| \hat{V}\hat{U}_{\tilde{t}}^\dagger | {\bf k +\delta \bf k} \big\rangle \right] \nonumber \\
=&\sum_{\bf p,{\bf p'} } \big \langle {V}_{{\bf p'},\bf k + \delta k}(0) {V}_{\bf k, \bf p}(\tilde{t})  \big\rangle\cdot \rho_{\bf p,\bf p'}(\tau) \cdot e^{i(E_{\bf k +\delta k} - E_{\bf p})\tilde{t}}. \\
\mathrm{III} 
= &\sum_{\bf p, {\bf p'} }  \mathrm{Tr}_{\mathrm{B}} \left[  \big \langle {\bf k}  | {U}_{\tilde{t}} {V} | {\bf p }   \big \rangle   \big \langle {\bf p} | \sigma(\tau) \rho_B^{eq} {U}_{\tilde{t}}^\dagger | {\bf p'}  \big\rangle  \big \langle {\bf p '}  | {V} | {\bf k +\delta k}  \big \rangle \right] \nonumber \\
= &\sum_{\bf p, {\bf p'} }  \big  \langle {V}_{\bf p',\bf k + \delta \bf k}(\tilde{t}) {V}_{\bf k,\bf p}(0)  \big \rangle \cdot \rho_{\bf p,\bf p'}(\tau) \cdot e^{i(E_{\bf p'} - E_{\bf k})\tilde{t}}. \\
\mathrm{IV} 
= &\sum_{\bf p, {\bf p'} }  \mathrm{Tr}_{\mathrm{B}} \left[  \big \langle {\bf k}  | {U}_{\tilde{t}} \sigma(\tau) \rho_B^{eq} | {\bf p }      \big\rangle  \big \langle {\bf p }   | {V} {U}_{\tilde{t}}^\dagger | {\bf p'}    \big \rangle  \big \langle {\bf p'}  | {V} | {\bf k +\delta \bf k}   \big  \rangle \right] \nonumber \\
= &\sum_{\bf p, {\bf p'} }   \big \langle {V}_{\bf p,\bf p '}(0) {V}_{\bf p', \bf k + \delta \bf k }(\tilde{t})  \big \rangle \cdot \rho_{\bf k,\bf p}(\tau) \cdot e^{i(E_{\bf p'} - E_{\bf k})\tilde{t}}. 
\end{align}

\subsection{Bath Correlation Functions}
The bath correlation function $\langle V_{\mathbf{k,p}}(\tilde{t}) \cdot V_{\mathbf{p,p'}}(0) \rangle$ in Eq.~\ref{term-I} can be written as 
\begin{align}
\Big\langle V_{\mathbf{k,p}}(\tilde{t}) \cdot V_{\mathbf{p,p'}}(0) \Big\rangle &= \gamma_{\mathbf{k,p}}\gamma_{\mathbf{p,p'}} \sum_j \frac{c_j^2}{2\omega_j} \Big\langle \big( b_{\mathbf{p-k},j}^\dagger(\tilde t) + b_{\mathbf{k-p},j}(\tilde t) \big) \big( b^\dagger_{\mathbf{p'-p},j}(0) + b_{\mathbf{p-p'},j}(0) \big) \Big\rangle \\
&= \gamma_{\mathbf{k,p}}\gamma_{\mathbf{p,p'}}  \sum_j \frac{c_j^2}{2\omega_j} \Big( \langle  b_{\mathbf{p-k},j}^\dagger(\tilde t)  b_{\mathbf{p-p'},j}(0)  \rangle + \langle b_{\mathbf{k-p},j}(\tilde t)  b^\dagger_{\mathbf{p'-p},j}(0)\rangle \Big).
\end{align}

where the terms in the form of $\langle b^\dagger b^\dagger \rangle$ and $\langle b b \rangle$ vanish upon taking the thermal expectation value. Note that, since $\langle  b_{\mathbf{p-k},j}^\dagger(\tilde t)  b_{\mathbf{p-p'},j}(0)  \rangle = \langle  b_{\mathbf{p-k},j}^\dagger(\tilde t)  b_{\mathbf{p-p'},j}(0)  \rangle \delta_{\mathbf{p-k},\mathbf{p-p'}}$ (or $\langle b_{\mathbf{k-p},j}(\tilde t)  b^\dagger_{\mathbf{p'-p},j}(0)\rangle = \langle b_{\mathbf{k-p},j}(\tilde t)  b^\dagger_{\mathbf{p'-p},j}(0)\rangle \delta_{ \mathbf{k -p},\mathbf{p'-p}}$), we get ${\bf k = p'}$. Consequently, we write this bath correlation function as 
\begin{align}
\Big\langle V_{\mathbf{k,p}}(\tilde{t}) \cdot V_{\mathbf{p,p'}}(0) \Big\rangle &= \gamma_{\mathbf{k,p}}\gamma_{\mathbf{p,p'}}\delta_{\mathbf{k,p'}} \sum_j \frac{c_j^2}{2\omega_j} \Big( n(\omega_j)e^{i\omega_j \tilde t} + [1 + n(\omega_j)]e^{-i\omega_j \tilde t} \Big)\nonumber\\
&= \gamma_{\mathbf{k,p}}\gamma_{\mathbf{p,p'}}\delta_{\mathbf{k,p'}}  \int_{0}^{\infty} d\omega J(\omega) \Big( n(\omega)e^{i\omega \tilde t} + [1 + n(\omega)]e^{-i\omega \tilde t} \Big)\nonumber\\
&= \gamma_{\mathbf{k,p}}\gamma_{\mathbf{p,p'}}\delta_{\mathbf{k,p'}}  \int_{-\infty}^{\infty} d\omega J(\omega)  n(\omega)e^{i\omega \tilde t}  = \gamma_{\mathbf{k,p}}\gamma_{\mathbf{p,p'}}\delta_{\mathbf{k,p'}}  \mathcal{B}(\tilde{t}). 
\end{align}
For the form of spectral density used here, 
$J(\omega) =  \sum_{j} \frac{a_j \omega }{[(\omega + \Omega_{j})^2 + \Lambda_{j}^2][(\omega - \Omega_{j})^2 + \Lambda_{j}^2]}$ we can obtain an analytical expression for $\mathcal{B}(\tilde{t})$ which is written as

\begin{align}
    \mathcal{B}(\tilde{t}) 
    &= \left[ \sum_j \frac{\pi a_j}{4\Omega_j \Lambda_j} \left( e^{i\Omega_j^+ \tilde t} n(\Omega_j^+) - e^{-i\Omega_j^- \tilde t} n(-\Omega_j^-) \right) + \frac{2\pi i}{\beta} \sum_{k=1}^{n'} J(i\mu_k)e^{-\mu_k \tilde t} \right] \equiv \sum_\alpha {\mathcal{B}}_\alpha e^{-\xi_\alpha \tilde{t}}.
\end{align}
Here, $\Omega_j^\pm = \Omega_j \pm i \Lambda_j$ and  $\mu_k = \frac{2\pi k}{\beta}$ are the Matsubara modes. In this work we have used $15$ Matsubara modes to obtain converged results. Note that we have now expressed $\mathcal{B}(\tilde{t})$ as a linear combination of exponential terms which we  use below to derive a time-local propagation scheme. Below we re-express the terms I-IV using  $\mathcal{B}(\tilde{t})$ as

\begin{align}\label{I-IV-final}
      \mathrm{I}
      &=\sum_{\bf p} \gamma_{\bf k,p}^2\cdot\mathcal{B}(\tilde{t}) \cdot \rho_{\bf k,k+\delta k}(\tau)e^{i (E_{\bf k+\delta k}- E_{\bf p}) \tilde{t}} \\
\mathrm{II} &=\sum_{\bf p}{\gamma_{\footnotesize \bf p + \delta \bf k,\bf k+\delta \bf k}\cdot \gamma_{\footnotesize \bf k,\bf p} \cdot \mathcal{B}^*(\tilde{t})}\cdot \rho_{\bf p,\bf p+\delta \bf k}(\tau)e^{i(E_{\bf k +\delta k} - E_{\bf p})\tilde{t}}\\
      \mathrm{III}&=\sum_{\bf p}  \gamma_{\footnotesize \bf p+\delta \bf k, \bf k+\delta \bf k} \cdot \gamma_{\bf k,\bf p }\cdot\mathcal{B}(\tilde{t})\cdot  \rho_{\footnotesize \bf p , \bf p+\delta \bf k}(\tau)e^{i (E_{ \bf p+\delta \bf k}- E_{\bf k}) \tilde{t}},\\
      \mathrm{IV}&= \sum_{\bf p}  \gamma^2_{\bf k + \delta \bf k,\bf p}\cdot\mathcal{B}^*(\tilde{t})\cdot\rho_{\bf k,k+\delta \bf k}(\tau)e^{i (E_{\bf p}- E_{\bf k}) \tilde{t}}.
\end{align}

\subsection{Exciton-Polariton Master Equation in the Interaction Picture}
Using the expression in Eq.~\ref{I-IV-final}, we find the following exciton-polariton master equation

\begin{align}\label{S45}
\dot{\rho}_{\bf k,k+\delta k}(t) &= - i (E_{\bf k} - E_{\bf k+\delta k})\rho_{\bf k,k+\delta k}(t)  \nonumber\\
& - \sum_{\alpha} \Bigg[ {\mathcal{B}}_\alpha\sum_{\bf p}  \gamma_{\bf \footnotesize k,p}^2\int_0^{t} d\tau \cdot \rho_{\bf k,k+\delta k}(\tau) \cdot  e^{i (E_{\bf k+\delta k}- E_{\bf p} + i \xi_{\alpha}) \tilde{t} } \Bigg]\nonumber\\
& +\sum_{\alpha} \Bigg[ {\mathcal{B}}^*_\alpha\sum_{\bf p}  \gamma_{\bf p+\delta k,k+\delta k} \cdot \gamma_{\bf k,p}\int_0^{t} d\tau \cdot \rho_{\bf p, p+\delta k}(\tau) \cdot  e^{i (E_{\bf k+\delta k}- E_{\bf p} + i \xi^*_{\alpha}) \tilde{t} } \Bigg] \nonumber\\
&+ \sum_{\alpha} \Bigg[ {\mathcal{B}}_\alpha\sum_{\bf p}  \gamma_{\bf p+\delta k,k+\delta k}\cdot\gamma_{\bf k,p}\int_0^{t} d\tau \cdot \rho_{\bf p,p+\delta k}(\tau) \cdot  e^{i (E_{\bf p+\delta k}- E_{ \bf k} + i \xi_{\alpha}) \tilde{t} } \Bigg] \nonumber\\
&- \sum_{\alpha} \Bigg[ {\mathcal{B}}^*_\alpha\sum_{\bf p}  \gamma_{\bf k+\delta k,p}^2\int_0^{t} d\tau \cdot \rho_{\bf k,k+\delta k}(\tau) \cdot  e^{i (E_{\bf p}- E_{\bf k} + i \xi^*_{\alpha}) \tilde{t} } \Bigg].
\end{align}

In the interaction picture, where $\rho^I_{k,k'}(t)=e^{i(E_k-E_{k'})t}\rho_{k,k'}(t)$, we write the master equation as
\begin{align}
\dot{\rho}^I_{\bf k, \bf k+\delta \bf k}(t) &= - \sum_{\alpha} \Bigg[ \mathcal{B}_\alpha  \sum_{\bf p }\gamma_{\bf k,\bf p}^2\int_0^{t} d\tau \cdot \rho^I_{\bf k,\bf k+\delta \bf k}(\tau) \cdot e^{\left[ i(E_{\bf k} - E_{\bf p}) - \xi_\alpha \right] t} e^{\left[ -i(E_{\bf k} - E_{\bf p }) + \xi_\alpha \right] \tau}\Bigg]\nonumber\\
&+ \sum_{\alpha} \Bigg[ \mathcal{B}^*_\alpha \sum_{\bf p} \gamma_{\bf p+ \delta \bf k, \bf k +\delta \bf k}\cdot \gamma_{\bf k,\bf p}\int_0^{t} d\tau \cdot \rho^I_{\bf p , \bf p +\delta k}(\tau) \cdot e^{\left[ i(E_{\bf k} - E_{\bf p}) - \xi^*_\alpha \right] t} \cdot e^{\left[ -i(E_{\bf k+\delta \bf k } - E_{\bf p +\delta \bf k}) + \xi^*_\alpha \right] \tau}\Bigg]\nonumber \\
&+ \sum_{\alpha} \Bigg[ \mathcal{B}_\alpha \sum_{\bf p} \gamma_{\bf p+\delta \bf k,\bf k+ \delta \bf k} \cdot\gamma_{\bf k,\bf p }\int_0^{t} d\tau \cdot \rho^I_{\bf p , \bf p +\delta \bf k}(\tau) \cdot e^{\left[ i(E_{\bf p + \delta \bf k} - E_{\bf k+\delta\bf  k}) - \xi_\alpha \right] t} \cdot e^{\left[ -i(E_{\bf p } - E_{\bf k}) + \xi_\alpha \right] \tau} \Bigg]\nonumber \\
&- \sum_{\alpha} \Bigg[ \mathcal{B}^*_\alpha  \sum_{\bf p}\gamma_{\bf k+\delta \bf k, \bf p}^2\int_0^{t} d\tau \cdot \rho^I_{\bf k,\bf k+\delta \bf k}(\tau) \cdot e^{\left[ i(E_{\bf p} - E_{\bf k+ \delta \bf k}) - \xi^*_\alpha \right] t} \cdot e^{\left[ -i(E_{\bf p} - E_{\bf k+\delta \bf k}) + \xi^*_\alpha \right] \tau} \Bigg].
\end{align}

Directly propagating this master equation, which involves a time-nonlocal propagation, is a numerically formidable task. Instead, 
we introduce four auxiliary time dependent 4-index tensors, $I_{\bf k, q}^{{\delta \bf k},\alpha}(t)$, $J_{\bf k, q}^{{\delta \bf k},\alpha}(t)$,  $F_{\bf k, q}^{{\delta \bf k},\alpha}(t)$ and $G_{\bf k, q}^{{\delta \bf k},\alpha}(t)$ to obtain the  following time-local equation of motion of $\dot{\rho}^I_{\bf k, k+\delta k}(t)$ as
\begin{align}
  \dot{\rho}^I_{\bf k, k+\delta k}(t) &=  - \sum_{\alpha,\bf p}\Big[ {\mathcal{B}}_\alpha  \gamma_{\bf p,k}^2\Big] \cdot I_{\bf k,\bf p}^{\delta \bf k,\alpha}(t)   + \sum_{\alpha,\bf p}\Big[ {\mathcal{B}}^*_\alpha  \gamma_{\bf k+\delta k,p+\delta k}
 \cdot \gamma_{\bf k,\bf p}\Big] \cdot F_{\bf k,\bf p}^{\delta \bf k,\alpha}(t) \nonumber \\
    &+\sum_{\alpha, \bf p}\Big[ {\mathcal{B}}_\alpha \gamma_{\bf k+\delta k,p+\delta k}\cdot  \gamma_{\bf k,p} \Big] \cdot G_{\bf k,\bf p}^{\delta \bf k,\alpha}(t)  -  \sum_{\alpha,\bf p}\Big[ {\mathcal{B}}^*_\alpha \gamma^2_{\bf k+\delta k,p}\Big] \cdot J_{\bf k,\bf p}^{\delta \bf k,\alpha}(t). 
\end{align}

\subsection{ Time-Local Evolution of Auxillary Tensors}
Here we derive a time-local propagation scheme for the four auxiliary tensors. Consider two complex scalar constants $A$ and $B$  representing the exponential factors in the bath correlation function. The general form of these  auxiliary tensors is written as
\begin{equation}\label{I(t)}
\mathcal{I}(t) = e^{At} \int_0^t d\tau \cdot \rho_{ab}(\tau) \cdot e^{B\tau}.
\end{equation}

The expression for  $\mathcal{I}(t + \Delta t)$  is written as
\begin{align}
\mathcal{I}(t + \Delta t) &= e^{A(t + \Delta t)} \int_0^{t + \Delta t} d\tau \cdot \rho_{ab}(\tau) \cdot e^{B\tau}  = e^{A\Delta t} e^{At} \left( \int_0^t d\tau \cdot \rho_{ab}(\tau) \cdot e^{B\tau} + \int_t^{t + \Delta t} d\tau \cdot \rho_{ab}(\tau) \cdot e^{B\tau} \right)
\nonumber\\
&= e^{A\Delta t} \left( \mathcal{I}(t) + e^{At} \int_t^{t + \Delta t} d\tau \cdot \rho_{ab}(\tau) \cdot e^{B\tau} \right) \approx e^{A\Delta t} \left( \mathcal{I}(t) + e^{At} \rho_{ab}(t) \int_t^{t + \Delta t} d\tau \cdot e^{B\tau} \right)\nonumber\\
&= e^{A\Delta t} \left( \mathcal{I}(t) + \rho_{ab}(t) e^{(A+B)t} \frac{e^{B\Delta t} - 1}{B} \right).
\end{align}

Using this expression, we obtain the following expression of the time-local propagation of the  auxiliary tensors as
\begin{align}
I_{\bf k,\bf q}^{\delta \bf k,\alpha}(t + \Delta t) &= e^{\left[ i(E_{\bf k} - E_{\bf q}) - \xi_\alpha\right] \Delta t} \left( I_{\bf k,\bf q}^{\delta {\bf k},\alpha}(t) + \frac{\rho^{I}_{\bf k, \bf k+\delta \bf k}(t)}{-i(E_{\bf k} - E_{\bf q})+ \xi_\alpha} \left[ e^{\left( -i(E_{\bf k} - E_{\bf q}) + \xi_\alpha \right) \Delta t} - 1 \right] \right), \nonumber \\
F_{\bf k,\bf q}^{\delta \bf k,\alpha}(t+\Delta t) &= e^{\left[ i(E_{\bf k} - E_{\bf q}) - \xi^{*}_\alpha \right] \Delta t} \left( F_{\bf k,\bf q}^{\delta \bf k,\alpha}(t)   + \frac{\rho^{I}_{\bf q, \bf q+\delta \bf k}(t) e^{i(E_{\bf k} - E_{\bf q} - E_{\bf k+\delta k} + E_{\bf q +\delta k})t}}{-i(E_{\bf k+\delta \bf k} - E_{\bf q +\delta \bf k}) + \xi^{*}_\alpha}   \left[e^{\left( -i(E_{\bf k+\delta \bf k} - E_{\bf q +\delta \bf k}) + \xi^{*}_\alpha \right) \Delta t}- 1\right]\right), \nonumber\\
G_{\bf k,\bf q}^{\delta \bf k,\alpha}(t+\Delta t) &= e^{\left[ i(E_{\bf q+\delta \bf k} - E_{\bf k+\delta \bf k}) - \xi_\alpha\right] \Delta t} \left( G_{\bf k,\bf q}^{\delta \bf k,\alpha}(t)   + \frac{\rho^{I}_{\bf q, \bf q +\delta \bf k}(t) e^{i(E_{\bf q +\delta \bf k} - E_{\bf k +\delta k} - E_{\bf q } + E_{\bf k})t}}{-i(E_{\bf q} - E_{\bf k}) + \xi_\alpha}   \left[e^{\left( -i(E_{\bf q} - E_{\bf k }) + \xi_\alpha \right) \Delta t}- 1\right]\right),\nonumber \\
J_{\bf k,\bf q}^{\delta \bf k,\alpha}(t + \Delta t) &= e^{\left[ i(E_{\bf q} - E_{\bf k +\delta \bf k}) - \xi^{*}_\alpha \right] \Delta t} \left( J_{\bf k,\bf q}^{\delta \bf k,\alpha}(t) + \frac{\rho^{I}_{\bf k, \bf k +\delta \bf k}(t)}{-i(E_{\bf q } - E_{\bf k +\delta \bf k}) + \xi^{*}_\alpha} \left[ e^{\left( -i(E_{\bf q} - E_{\bf k +\delta \bf k}) + \xi^{*}_\alpha \right) \Delta t} - 1 \right] \right).
\end{align}

\subsection{Coarse grained master equation}
To efficiently propagate the exciton-polariton dynamics we adapt an asymmetric sampling of the momentum space, such that the master equation is written as

\begin{align}\label{ME}
  \dot{\rho}^I_{\bf k, k+\delta k}(t) &= -\sum_{ {\bf q}\in \mathrm{CG}} n_{\bf q} \sum_{\alpha}\Big[ {\mathcal{B}}_\alpha  \gamma_{{\bf q},{\bf k}} ^2\Big] \cdot I_{\bf k,\bf q}^{\delta \bf k,\alpha}(t)   + \sum_{ {\bf q}\in \mathrm{CG}}n_{\bf q}   \sum_{\alpha}\Big[ {\mathcal{B}}^*_\alpha\gamma_{{\bf k}+\delta {\bf k},{\bf q}+\delta {\bf k}}
  \gamma_{{\bf k},{\bf q}}\Big] \cdot F_{\bf k,\bf q}^{\delta \bf k,\alpha}(t) \nonumber \\
    &+\sum_{ {\bf q}\in \mathrm{CG}}n_{\bf q} \sum_{\alpha}\Big[ {\mathcal{B}}_\alpha  \gamma_{{\bf k}+\delta {\bf k},{\bf q}+\delta {\bf k}}  \gamma_{{\bf k},{\bf q}} \Big] \cdot G_{\bf k,\bf q}^{\bf \delta  k,\alpha}(t)  - \sum_{ {\bf q}\in \mathrm{CG}}n_{\bf q}  \sum_{\alpha}\Big[ {\mathcal{B}}^*_\alpha\gamma^2_{{\bf k}+\delta {\bf k},{\bf q}}\Big] \cdot J_{\bf k,\bf q}^{\delta \bf k,\alpha}(t), 
\end{align}

where ${\bf q}\in \mathrm{CG}$, with $\mathrm{CG}$ representing a set of coarse grained (or sampled) points and $n_{\bf q} $ as the weight of each coarse-grained point ${\bf q}$. Specifically, we partition the Brillouin zone into three disjoint sub-domains,  where $\mathrm{CG} = \mathrm{CG}_\mathrm{I} ~\cup~ \mathrm{CG}_\mathrm{II}~\cup~\mathrm{CG}_\mathrm{III}$.  Region I, represented by $\mathrm{CG}_\mathrm{I}$,  includes the relatively flat regions of the dispersion near the Brillouin zone boundaries, and thus is heavily coarse-grained by sparsely sampling the wavevector space. We sample the region I with $N_I \times N_I - 1$ wavevector points ($N_I = 19$ chosen in this work), with the $\Gamma$ point excluded. A small fraction of the full Brillouin zone, centering the $\Gamma$ point, specified by a zoning factor $Z_\mathrm{F}$ (set  0.014 in this work) and spanning $-Z_\mathrm{F} \frac{\pi}{a} \leq k_x, k_y \leq Z_\mathrm{F}\frac{\pi}{a}$, is sampled more densely; we denote this as Region $\mathrm{II}$. Finally, Region $\mathrm{III}$, which defines the specific excitation window used to initialize the exciton-polariton density, is retained at the full wavevector resolution without any coarse-graining. Region $\mathrm{II}$ is initially discretized using an $N_{\mathrm{II}}\times N_{\mathrm{II}}$ (with $N_{II} = 35$ chosen in this work) uniform grid , after which all grid points overlapping with the excitation window (Region $\mathrm{III}$) are removed and replaced by the corresponding fine-grid points. To preserve the total statistical weight, the weight associated with the newly introduced Region $\mathrm{III}$ points is compensated by adjusting the weights of the four nearest grid points in the Region $\mathrm{II}$. This excitation window is defined as a ring in momentum space, which contains all the wavevectors within a radius of $k_r \cdot\delta k$ from an excitation centered at ${\bf k}_0$. Here, $k_r$ is set to $2.5$ and $\delta k = \frac{2 \pi}{L} = \frac{2 \pi}{\sqrt{N}\cdot a} $.

\subsection{Markovian Master Equations}
Below we derive the Markovian master equation for propagating the exciton-polariton density. Specifically we start from Eq.~\ref{S45}, and impose the Markovian approximation, i.e., $\int_0 ^t \mathcal{K}(t-\tau)\hat{\rho}(\tau)d\tau \approx \hat{\rho}(t)\int_0 ^\infty \mathcal{K}(\tau)d\tau$, to obtain the following Markovian master equation

\begin{align}
\dot{\rho}_{\bf k,k+\delta k}(t) &\approx - i (E_{\bf k} - E_{\bf k+\delta k})\rho_{\bf k,k+\delta k}(t)  \nonumber\\
& - \sum_{\alpha} \Bigg[ {\mathcal{B}}_\alpha\sum_{\bf p}  \gamma_{\bf \footnotesize k,p}^2 \cdot \rho_{\bf k,k+\delta k}(t)\cdot \int_0^{\infty} dt~     e^{i (E_{\bf k+\delta k}- E_{\bf p} + i \xi_{\alpha}) t } \Bigg]\nonumber\\
& +\sum_{\alpha} \Bigg[ {\mathcal{B}}^*_\alpha\sum_{\bf p}  \gamma_{\bf p+\delta k,k+\delta k} \gamma_{\bf k,p} \cdot  \rho_{\bf p, p+\delta k}(t) \cdot \int_0^{\infty} dt  ~ e^{i (E_{\bf k+\delta k}- E_{\bf p} + i \xi^*_{\alpha}) t} \Bigg] \nonumber\\
&+ \sum_{\alpha} \Bigg[ {\mathcal{B}}_\alpha\sum_{\bf p}  \gamma_{\bf p+\delta k,k+\delta k}\gamma_{\bf k,p} \cdot \rho_{\bf p,p+\delta k}(t) \cdot \int_0^{\infty} dt ~  e^{i (E_{\bf p+\delta k}- E_{ \bf k} + i \xi_{\alpha}) t } \Bigg] \nonumber\\
&- \sum_{\alpha} \Bigg[ {\mathcal{B}}^*_\alpha\sum_{\bf p}  \gamma_{\bf k+\delta k,p}^2\cdot \rho_{\bf k,k+\delta k}(t) \cdot\int_0^{\infty} dt ~  e^{i (E_{\bf p}- E_{\bf k} + i \xi^*_{\alpha}) t } \Bigg].\\
 &\approx - i (E_{\bf k} - E_{\bf k+\delta k})\rho_{\bf k,k+\delta k}(t)  \nonumber\\
& - \sum_{\alpha} \Bigg[ {\mathcal{B}}_\alpha\sum_{\bf p}  \gamma_{\bf \footnotesize k,p}^2\cdot \rho_{\bf k,k+\delta k}(t)\cdot \frac{1}{\xi_{\alpha} -i (E_{\bf k+\delta k}- E_{\bf p})} \Bigg]\nonumber\\
& +\sum_{\alpha} \Bigg[ {\mathcal{B}}^*_\alpha\sum_{\bf p}  \gamma_{\bf p+\delta k,k+\delta k} \gamma_{\bf k,p} \cdot  \rho_{\bf p, p+\delta k}(t) \cdot \frac{1}{\xi^*_{\alpha} -i (E_{\bf k+\delta k}- E_{\bf p} )} \Bigg] \nonumber\\
&+ \sum_{\alpha} \Bigg[ {\mathcal{B}}_\alpha\sum_{\bf p}  \gamma_{\bf p+\delta k,k+\delta k}\gamma_{\bf k,p} \cdot \rho_{\bf p,p+\delta k}(t) \cdot \frac{1}{\xi_{\alpha}-i (E_{\bf p+\delta k}- E_{ \bf k})}\Bigg] \nonumber\\
&- \sum_{\alpha} \Bigg[ {\mathcal{B}}^*_\alpha\sum_{\bf p}  \gamma_{\bf k+\delta k,p}^2\cdot \rho_{\bf k,k+\delta k}(t) \cdot \frac{1}{ \xi^*_{\alpha}- i (E_{\bf p}- E_{\bf k})} \Bigg].
\end{align}
The corresponding Markovian master equation in the interaction picture, with
${\rho}_{\bf k,k+\delta k}^{{I}}(t) = e^{i\Delta E_{\bf k, k+\delta k}t}{\rho}_{\bf k,k+\delta k}(t)$, is written as
\begin{align}
\dot{\rho}^{{I}}_{\bf k,k+\delta k}(t) =
& - \sum_{\alpha} \Bigg[ {\mathcal{B}}_\alpha\sum_{\bf p}  \gamma_{\bf \footnotesize k,p}^2\cdot {\rho}^{{I}}_{\bf k,k+\delta k}(t)\cdot \frac{1}{\xi_{\alpha} -i (E_{\bf k+\delta k}- E_{\bf p})} \Bigg]\nonumber\\
& +\sum_{\alpha} \Bigg[ {\mathcal{B}}^*_\alpha\sum_{\bf p}  \gamma_{\bf p+\delta k,k+\delta k} \gamma_{\bf k,p} \cdot  {\rho}^{{I}}_{\bf p, p+\delta k}(t) \cdot \frac{e^{i\Delta E_{\bf k, k +\delta k}t} e^{-i\Delta E_{\bf p, p +\delta k}t}}{\xi^*_{\alpha} -i (E_{\bf k+\delta k}- E_{\bf p} )} \Bigg] \nonumber\\
&+ \sum_{\alpha} \Bigg[ {\mathcal{B}}_\alpha\sum_{\bf p}  \gamma_{\bf p+\delta k,k+\delta k}\gamma_{\bf k,p} \cdot \rho^{{I}}_{\bf p,p+\delta k}(t) \cdot \frac{e^{i\Delta E_{\bf k, k +\delta k}t} e^{-i\Delta E_{\bf p, p +\delta k}t}}{\xi_{\alpha}-i (E_{\bf p+\delta k}- E_{ \bf k})}\Bigg] \nonumber\\
&- \sum_{\alpha} \Bigg[ {\mathcal{B}}^*_\alpha\sum_{\bf p}  \gamma_{\bf k+\delta k,p}^2\cdot {\rho}^{{I}}_{\bf k,k+\delta k}(t) \frac{1}{ \xi^*_{\alpha}- i (E_{\bf p}- E_{\bf k})} \Bigg].
\end{align}

Therefore, time-dependent auxiliary tensors under the Markovian limit is written as
\begin{align}
{I}_{\bf k,q}^{{\bf \delta k},\alpha}(t) =  \frac{\rho^I_{\bf k, k+\delta k}(t)}{-i(E_{\bf k+\delta k} - E_{\bf q}) + \xi_\alpha},  
~~~~&{F}_{\bf k,q}^{{\bf \delta k},\alpha}(t) =  \frac{\rho^I_{\bf q, q+\delta k}(t)\cdot e^{i(E_{\bf k} - E_{\bf q} - E_{\bf k+\delta k} + E_{\bf q +\delta k})t}}{-i(E_{\bf k+\delta k} - E_{\bf q}) + \xi^*_\alpha}, \\
{G}_{\bf k,q}^{{\bf \delta k},\alpha}(t) =  \frac{\rho^I_{\bf q, q+\delta k}(t)\cdot e^{i(E_{\bf k} - E_{\bf q} - E_{\bf k+\delta k} + E_{\bf q +\delta k})t}}{-i(E_{\bf q+\delta k} - E_{\bf k}) + \xi_\alpha},  ~~~~
&{J}_{\bf k,q}^{{\bf \delta k},\alpha}(t) =  \frac{\rho^I_{\bf k, k+\delta k}(t) }{-i(E_{\bf q} - E_{\bf k}) + \xi^*_\alpha}. 
\end{align}

\section{Extraction of molecular spectral density}
We extract the spectral density describing the exciton–phonon coupling from direct molecular dynamics simulations using DFTB by computing the time-dependent energy-gap fluctuations with the TD-DFTB approach.~\cite{gustin2023mapping, valleau2012alternatives, plotz2017spectral,makri1998quantum}  This approach relies on mapping a molecular system to a two level system coupled to set of harmonic modes, such that the ground  and excited state energies are expressed as

\begin{align}
E_g &= E_g^0 + \sum_j \frac{1}{2}   \omega_j^2 R_j^2, \\
E_e &= E_e^0 + \sum_j \frac{1}{2}  \omega_j^2 R_j^2 + \sum_j c_j R_j.
\end{align}

The energy gap between the excited and ground state energies is given as
\begin{align}
\Delta E(t) &= E_e - E_g = \Delta E^0 + \sum_j c_j R_j(t),
\end{align}
where $\Delta E^0 = E_e^0 - E_g^0$. The time-dependent nuclear motion during ground-state molecular dynamics is given by
\begin{align}
R_j(t) &= R_j(0)\cos(\omega_j t) + \frac{P_j(0)}{\omega_j}\sin(\omega_j t).
\end{align}

Since $\langle P_j(0)\rangle=0$ and $\langle R_j(0) \rangle=0$ (where $\langle..\rangle$ denotes ensemble averaging over thermally sampled trajectories), the mean energy gap is the static gap $\langle\Delta E(t)\rangle = \Delta E^0$. We define the energy fluctuation $\bar{E}(t)$ as follows,
\begin{align}
\bar{E}(t) &= \Delta E(t) - \langle\Delta E\rangle = \sum_j c_j \left( R_j(0)\cos(\omega_j t) + \frac{P_j(0)}{  \omega_j}\sin(\omega_j t) \right).
\end{align}
To extract the spectral density, we evaluate the energy gap auto-correlation function $\langle\bar{E}(t)\bar{E}(0)\rangle$. The cross terms and sine terms vanish upon averaging, giving the following equations,
\begin{align}\label{ecorr}
\langle\bar{E}(t)\bar{E}(0)\rangle &= \sum_j c_j^2 \langle R_j^2(0)\rangle \cos(\omega_j t) = \frac{1}{
\beta} \sum_j \frac{c_j^2  }{  \omega_j^2} \cos(\omega_j t).
\end{align}

Note that $\langle\bar{E}(0)\bar{E}(0)\rangle = \frac{2}{\beta} \int_{0}^{\infty} d\omega \frac{1}{\omega}J(\omega) = \frac{2}{\beta} \lambda_s$,  where $\lambda_s$ is the reorganization energy. The spectral density is then obtained from the energy-gap autocorrelation function through a discrete cosine transform,
\begin{align}
J(\omega)
=
\frac{\beta\omega}{\pi}
\int_0^\infty dt\,
\left\langle \bar{E}(t)\bar{E}(0)\right\rangle
\cos(\omega t).
\end{align}

\subsection*{Parameters obtained from DFTB/TD-DFTB Simulations}

We obtain the molecular parameters following the strategy outlined in the previous section using outputs from molecular dynamics simulation of DFTB/TD-DFTB calculations of TIPS pentacene (we consider a $2\times2\times1$ supercell). We generate 50 initial geometries from one DFTB-MD (NVT) trajectory, with each geometry assigned a random velocity sampled from a thermal Boltzmann distribution at  $T=300$ K. We used the ‘SupercellFolding’ scheme to perform on-the-fly MD with SCC-DFTB on a 3$\times$3$\times$3 folded supercell, where we used $\Gamma$-point sampling to approximate Brillouin zone integration. Along these 50 trajectories, we perform  TD-DFTB calculations to extract the energy gap fluctuation using the $\textsc{pbc-0-3}$ Slater-Koster parameter set. Note that the unit cell used in this work contains two TIPS-pentacene molecules, we extract the geometry of one molecule to compute the time-dependent localized excitonic energy. The extracted spectral density is then fitted using the following analytical model

\begin{align}
J(\omega) &= \sum_{{\alpha}=1}^5 \frac{A_{\alpha} \omega}{\big((\omega+\Omega_{\alpha})^2+\Gamma_{\alpha}^2\big)\big((\omega-\Omega_{\alpha})^2+\Gamma_{\alpha}^2\big)},
\end{align}
where $\Omega_{\alpha}$ is the central oscillator frequency, $\Gamma_{\alpha}$ is the damping rate, and $A_{\alpha}$ defines the coupling strength. The parameters obtained via this procedure are presented in Table~\ref{t1} (using the atomic units). The rest of the parameters used in our quantum dynamical simulations are presented in Table~\ref{t2}.

\begin{table}[htpb]
\centering
\caption{Parameters for fitting the TD-DFTB spectral density}
\label{tab:bath_parameters}
\begin{tabular}{lc} 
\hline\hline
Parameter  & Values (a.u.) \\
\hline
$\Omega_{\alpha}$  & \{0.00041, 0.00703, 0.00128, 0.00785, 0.0059\} \\ 
$\Gamma_{\alpha}$ & \{0.0003, 0.00017, 0.0003, 0.00018, 0.0002\} \\
$A_{\alpha} (\times 10^{-11})$       & \{0.006511, 3.52945, 0.035967, 0.72834, 0.3597 \} \\ 
\hline\hline\label{t1}
\end{tabular}
\end{table}


\begin{table}[htpb]
\centering
\caption{System and numerical simulation parameters for TIPS Pentacene crystal exciton-polariton dynamics.}
\label{tab:parameters}
\begin{tabular}{lcc}
\hline\hline
Parameter & Symbol & Value \\
\hline
Material's refractive index & $\eta$ & 1.8 \\
Lattice constant & $a$ & 15 \AA \\
Number of lattice sites & $N$ & 12001$\times$12001 \\
Excitonic energy & $\epsilon_{0}$ & 1.96 eV \\
Cavity frequency & $\omega_{c}(0)$ & 1.30 eV \\
Light-matter coupling & $g_c$ & 0.15 eV \\
Inverse temperature ($T = 300 K$) & $\beta$ & 1052.8 a.u. \\
Time step & $\Delta t$ & 100.0 a.u. \\
\hline\hline\label{t2}
\end{tabular}
\end{table}

\section{Calculation of Group Velocity Renormalization}
To quantify the effect of the phonon bath on the transport properties of the system, we calculate the percentage change in the group velocity,  given by the following relation of the group velocity renormalization $\mathcal{V}_g$ percentage,
\begin{equation}
  \mathcal{V}_g = \left( \frac{v_g}{v_g^0} - 1 \right) \times 100
\end{equation}
Here, $v_g^0$ represents the bare group velocity of the exciton-polariton wavepacket in the absence of phonon coupling, and $v_g$ is the group velocity in the presence of the coupling to the phonon bath. Both velocities are extracted by tracking the trajectory of the maximum of the real space probability density of the wavepacket over time.

\section{On fitting experimental data to a 2-band polariton model} 
\begin{figure*}
\centering
\includegraphics[width=1.0\linewidth]{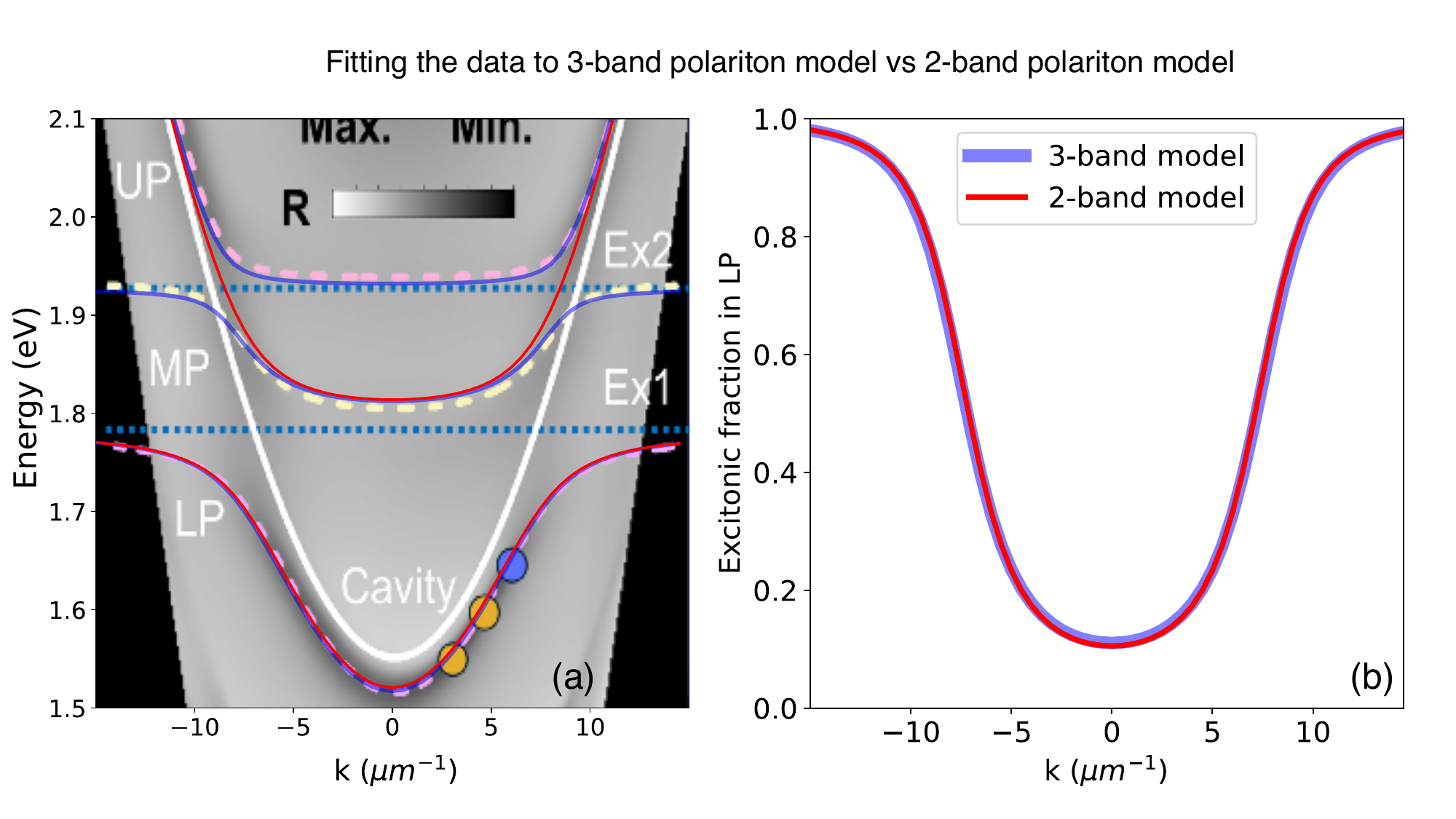}
\caption{\footnotesize (a) Angle-resolved reflectance spectra of a cavity filled with TIPS-pentacene crystal (adapted with permission from Fig.~\cite{hong2026exciton}; Copyright 2026 Elsevier) showing reasonable fits to a 3-band (blue) and a simplified 2-band (red) dispersion model. (b) Excitonic fraction in the LP versus k. Energy is reported in eV, while the magnitude of wave-vector ($k$) is reported in $\mu m^{-1}$.}
\label{fig1}
\end{figure*}

Fig. 2 of Ref.~\cite{hong2026exciton}, which provides experimental  group velocities and their renormalization in TIPS-pentacene  is compared to  the predictions of our quantum dynamical simulations in the main-text. In  this experimental work~\cite{hong2026exciton}, the polaritonic angle-resolved reflectance spectra (illustrated in Fig~\ref{fig1}a) was fitted to a 3-band polariton model (blue curves). We find that the  lower polariton band can also be  fitted  using a simpler two-band  dispersion model (red curve) that yields identical group velocities and more importantly visually identical excitonic character, as illustrated in Fig~\ref{fig1}a-b.  

\section{Incoherent Dynamics of Excitons}
\begin{figure*}
\centering
\includegraphics[width=1.0\linewidth]{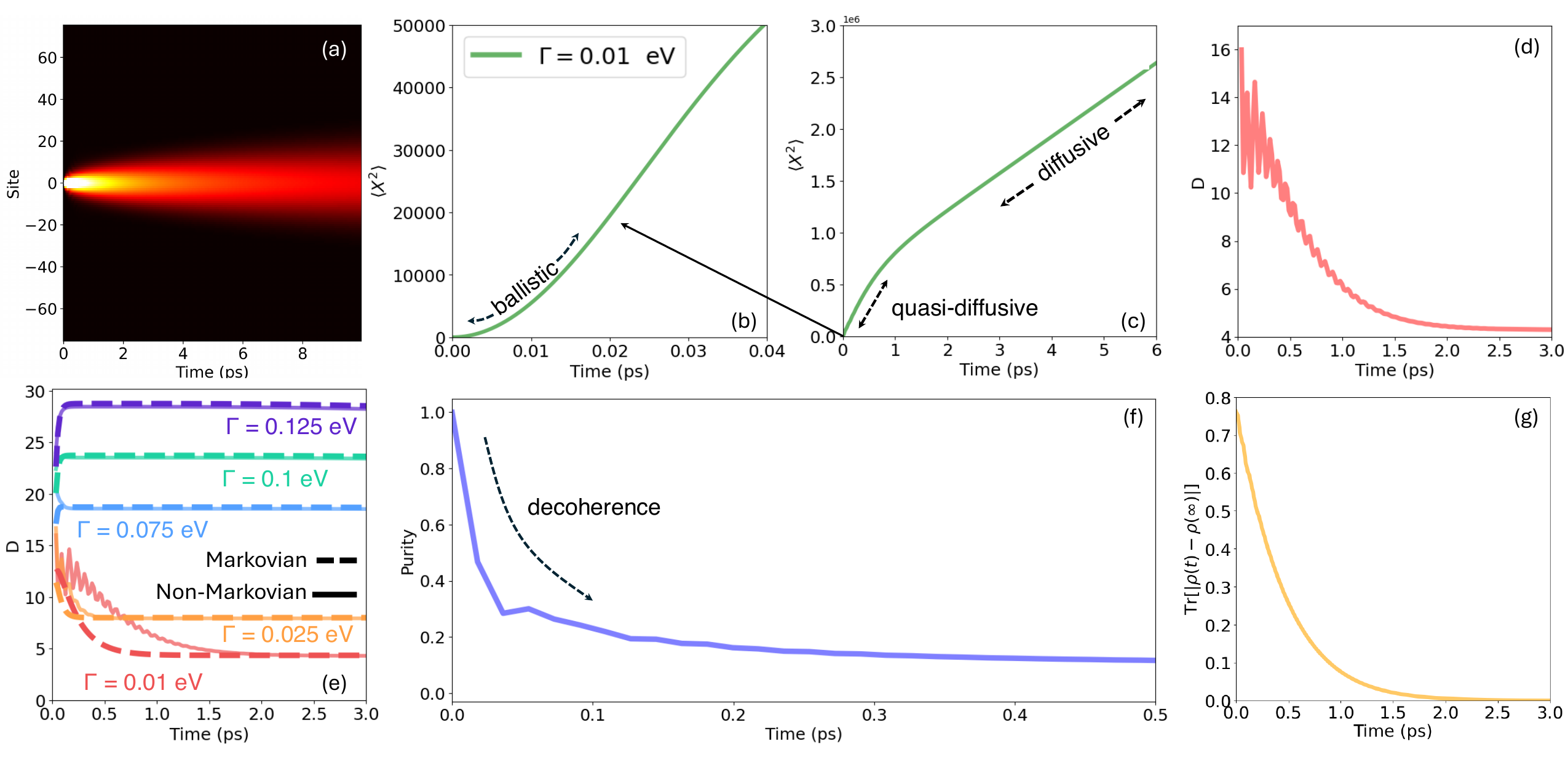}
\caption{\footnotesize (a) Time-dependent propagation of a centrally initialized and normalized excitonic density over a lattice having 151 sites. (b)-(c) Early-time ballistic (b) and thereafter, diffusive excitonic transport (c) illustrated through plotting of the mean squared displacement (MSD) over time (in ps). (d) Variation of the diffusion constant (D) over time (in ps). (e) Time-dependent diffusion constants (D) plotted for different widths of the bath spectral density ($\hbar\Gamma=0.01,0.025,0.075,0.1,0.125$ eV). (f) Decoherence observed through the decay of the excitonic purity over time. (g) Thermalization of the excitonic population, illustrated by a plot of $Tr[|\rho(t)-\rho(\infty)|]$ versus time. The width of the spectral density associated with the phonon bath is $\hbar\Gamma=0.01$ eV, for (a)-(d) and (f)-(g).}
\label{fig2}
\end{figure*}

In this section we discuss the incoherent diffusive transport of excitons and exciton-polaritons. The diffusion constant under the Markovian approximation is written as~\cite{10.1063/1.434789, Munn01011980} 
\begin{align}\label{Silbey-Diff}
D \approx \sum_k \frac{v_{\bf k}^2}{\Gamma_{\bf kk}}\rho_{\bf kk}^{eq}
\end{align} 

where $\rho_{\bf kk}^{eq}$ is the population of state ${\bf k}$ at equilibrium and $\Gamma_{\bf kk} = \sum_{\bf p} \int_0^\infty dt \big[ \langle \hat{V}_{\bf k,p} (t) \hat{V}_{\bf p,k}    \rangle \cdot e^{i(E_{\bf k}-E_{\bf p})t}  +  \langle \hat{V}_{\bf k,p}   \hat{V}_{\bf p,k}(t)   \rangle e^{i(E_{\bf p}-E_{\bf k})t}\big]$ is the phonon scattering rate. Importantly, such a treatment has two shortcomings: (i) the observed diffusion constant can deviate significantly from this estimate, if the reduced density matrix has not yet thermalized, and (ii) the Markovian approximation itself may break down. We illustrate these two aspects in Fig~\ref{fig2}, which presents the incoherent dynamics of excitons over a 1D lattice with $N = 151$ sites with the corresponding Hamiltonian written as 
\begin{align}\label{EX-PH}
\hat{{H}}_{\mathrm{ex}} 
&= \sum_{ k} \hat{X}_{ k}^{\dagger}\hat{X}_{ k} E_k  + {\sum_{{ k}, j}  \hat{b}^{\dagger}_{{ k},j}\hat{b}_{{ k},j}\omega_{j}}  +  \sum_{{ k},{ \delta k}}  \hat{X}_{ k+\delta k}^{\dagger}\hat{X}_{{ k}} \Big[\sum_j \frac{c_j}{\sqrt{2N\omega_j}} (\hat{b}_{-\delta { k}, j}^{\dagger} + \hat{b}_{\delta { k},j})\Big],
\end{align}
where $E_k  = - 2\tau \cos(k\cdot a) $ with $\tau = 150 ~\mathrm{cm}^{-1}$ and $a=100~\mathrm{\AA}$. Here we sample $\{\omega_j, c_j\}$ from the spectral density $J(\omega)=\frac{A\omega}{{((\omega+\Omega)^2+\Gamma^2)((\omega-\Omega)^2+\Gamma^2)}}$ with $A = 4\Gamma   (\Gamma^2 + \Omega^2) \frac{\lambda_s}{\pi} $ with $\lambda_s = 0.000746$ a.u. and $\Omega = 2$ meV. Fig~\ref{fig2}a shows the time-dependent excitonic (spatial) density which exhibits incoherent (hence diffusive) transport.  The mean-square displacement of this excitonic density is presented  Fig~\ref{fig2}b-c which feature three distinct regimes of transport: a short time ballistic regime, an intermediate quasi-diffusive regime and an equilibrium diffusive regime. Here we compute the mean squared displacement (MSD) from the simulations as
\begin{align}
    \langle X^2(t)\rangle= -\lim_{\bf \kappa\rightarrow 0}\nabla^2_{\bf \kappa}\sum_{\bf \kappa}\rho_{\bf k,k+\kappa}(t),
\end{align}
 where we evaluate the second-derivative with respect to $\kappa$ using the $8^{th}$ order central difference approach. Our numerical results show that a ballistic transport persists for $\sim 10$ fs which is corroborated in Fig.~\ref{fig2}f that also displays the decay of purity in the same time-scale. Once the system becomes incoherent it shows a quasi-diffusive transport with its MSD approximately staying linear for the next $\sim 0.5$ ps. This diffusive transport does not correspond to the typical thermalized diffusion described using Eq.~\ref{Silbey-Diff}. This is because even though the reduced density matrix has decohered it has not reached equilibrium and as a result this regime corresponds to a non-equilibrium (i.e. hot) diffusion. As the density matrix equilibrates the diffusion constant reaches its equilibrium value and consequently the MSD grows linearly at longer times. This dynamical evolution of the diffusion constant due to the time-dependent relaxation of the density can be observed in  Fig~\ref{fig2}d which displays the time-dependent diffusion constant. Fig~\ref{fig2}g shows the quantum distance between the instantaneous  density matrix and its equilibrium counterpart and exhibits the same trend as in  Fig~\ref{fig2}d confirming our assertion on the nature of the excitonic incoherent dynamics. 

It is important to note that the incoherent quasi-diffusive regime is not necessarily Markovian. For a spectral density featuring sharp peaks (as is the case for TIPS-pentacene), the Markovian treatment breaks down. This is illustrated in Fig.~\ref{fig2}e which presents the time-dependent diffusion constants at different values of $\Gamma$ which controls the width of the spectral density obtained from the Markovian and the non-Markovian quantum dynamical simulations. At lower $\Gamma$, the differences between the Markovian and the non-Markovian exciton dynamics are pronounced, as is expected from a sharp spectral density which causes a slower decay in the bath correlation function over time. In contrast, at higher values of  $\Gamma$, a broadening of the spectral density allows much faster damping of the correlation function, and as a result, the Markovian treatment reproduces the non-Markovian predictions. This means, in the context of polaritons, when the diffusive regime sets in after ballistic propagation, typically in sub-picosecond timescale, it is likely that the exciton-polariton density remains highly non-equilibrium and non-Markovian. As a result, the typical diffusive regime observed in related experiments should not be studied using Markovian treatment or the equilibrium expression for the diffusion constant in Eq.~\ref{Silbey-Diff}.

\clearpage
\bibliography{bib}